\documentclass[aps,prd,preprint,nofootinbib,11pt]{revtex4}
\usepackage{amsfonts}
\usepackage{mathrsfs}
\usepackage{graphicx}
\usepackage{amsmath}
\usepackage{amssymb}
\usepackage{multirow}
\usepackage{subfigure}
\usepackage{epsfig}
\usepackage{color}
\usepackage{booktabs}
\usepackage[section]{placeins}
\usepackage[colorlinks=true,linkcolor=blue,urlcolor=blue,citecolor=blue]{hyperref}


\providecommand{\doi}[1]{\href{https://doi.org/#1}{\nolinkurl{#1}}}
\begin{document}

\title{QCD Sum Rule Analysis of a Compact $D^{+}D^{-}K^{+}$-Like Hidden-Charm Hexaquark with $J^{P}=0^{-}$\\[0.7cm]}

\author{Jing-Yi Yan}
\author{Wen-Shuai Zhang}
\author{Liang Tang}%
 \email{tangl@hebtu.edu.cn}
\affiliation{%
\textit{College of Physics and Hebei Key Laboratory of Photophysics Research and Application, }
\\
\textit{Hebei Normal University, Shijiazhuang 050024, China} %\textbackslash\textbackslash
}%

%\date{\today}% It is always \today, today,
             %  but any date may be explicitly specified
%\affiliation{}

\begin{abstract}
\vspace{0.3cm}
In this work, we study a compact hexaquark configuration motivated by the same quark content as $D^{+} D^{-} K^+$ using QCD sum rules, where the $D D K$ system has been extensively studied within theoretical frameworks of few-body hadronic dynamics and coupled-channel interactions. The state is constructed from three color-octet quark--antiquark clusters coupled to an overall color singlet. We construct six independent local interpolating currents with the quantum numbers $J^{P}=0^{-}$ and analyze the corresponding two-point correlation functions. Both perturbative contributions and nonperturbative condensates up to dimension ten are included in the operator product expansion. Our analysis indicates that, within Borel windows satisfying standard sum rule criteria and with a reasonable choice of continuum threshold, the mass of the $J^{P}=0^{-}$ hidden-charm hexaquark state is estimated to be in the range $3.94$--$4.41~\mathrm{GeV}$. This prediction can provide a valuable theoretical reference for identifying such exotic hidden-charm hexaquark state in future experiments.

\end{abstract}
\maketitle
\newpage

\section{Introduction}

Quantum chromodynamics (QCD) is widely accepted as the fundamental theory describing the strong interaction among quarks and gluons. In the conventional quark model proposed by Gell-Mann and Zweig~\cite{Gell-Mann:1964ewy,Zweig:1964ruk}, hadrons are classified as mesons, composed of a quark--antiquark pair (\(q\bar{q}\)), and baryons, composed of three quarks (\(qqq\)). However, the underlying principles of QCD do not restrict hadronic states to these minimal configurations. More complex color-singlet combinations, such as tetraquarks, pentaquarks, and hexaquark systems (including both \(qqqqqq\) and \(qqq\bar{q}\bar{q}\bar{q}\) configurations), are allowed and have long been anticipated~\cite{Jaffe:1976yi,Chen:2016qju,Lebed:2016hpi,WASA-at-COSY:2014dmv}. The exploration of such exotic hadrons provides a unique opportunity to probe the nonperturbative dynamics of QCD, including color confinement and multiquark correlations.

For more than two decades, hadron spectroscopy has entered a new era driven
by high-statistics data from $e^+e^-$ collider experiments (Belle/Belle~II,
BaBar, BESIII), hadron collider experiments (CDF, D0, LHCb, CMS), and fixed-target experiments (COMPASS, GlueX), which together have
revealed a rich spectrum of exotic and conventional hadron states. Since the discovery of the \(X(3872)\) by the Belle Collaboration in 2003~\cite{Belle:2003nnu}, a large number of unconventional hadronic states, commonly referred to as the \(XYZ\) states, as well as the \(P_c\) pentaquarks, have been observed~\cite{LHCb:2016lve,LHCb:2019kea,Ali:2017jda,Esposito:2016noz,Richard:2016eis,Liu:2019zoy,Brambilla:2019esw,BESIII:2013ris,Wang:2019tlw}. Many of these states cannot be accommodated within the conventional quark model, stimulating extensive theoretical interpretations. Among the most widely discussed scenarios are hadronic molecules formed by loosely bound color-singlet hadrons~\cite{Guo:2017jvc,Voloshin:2004mh,Tornqvist:2004qy}, compact multiquark configurations driven by diquark correlations~\cite{Maiani:2004vq,Ali:2017jda,Wang:2013vex}, and kinematic effects associated with thresholds~\cite{Guo:2015umn,Szczepaniak:2015eza}. The hexaquark states investigated in the literature are predominantly based on dibaryon or baryonium configurations. For instance, the doubly heavy hexaquark states studied in Refs.~\cite{Wang:2017sto,
Wang:2021qmn,
Wan:2019ake,
Zhang:2025jqx,
Wan:2023epq,
Wang:2026lta,
Chen:2016ymy,Wang:2021pua} have yielded physically meaningful results. Despite significant progress, a coherent and universally accepted picture of exotic hadrons has not yet emerged.

A natural extension of these studies is the investigation of systems with even richer structures, in particular three-body hadronic systems. Analogous to few-body systems in nuclear and atomic physics, where multi-particle binding can arise from pairwise interactions, it is natural to explore whether three hadrons can form bound or resonant states under sufficiently strong attractive forces. In this context, systems composed of open-charm mesons have attracted increasing attention. In particular, \(DDK\)-type systems have been studied using effective field theory, unitarized approaches, and Faddeev-type equations~\cite{MartinezTorres:2019,PhysRevD.103.L031501,Wu:2019vsy,Wu:2025fzx,Huang:2019qmw,Zhang:2024yfj}. These studies indicate that the strong attraction in the \(DK\) subsystem---closely related to the dynamics responsible for the \(D_{s0}^*(2317)\)---may support weakly bound or resonant three-body configurations, highlighting the open-charm sector as a promising arena for few-body dynamics.

It should be emphasized, however, that such approaches describe spatially extended, nonlocal hadronic molecules. An important and largely unexplored question is whether the same underlying quark content, schematically \(c\bar{d}\,d\bar{c}\,u\bar{s}\), can also give rise to alternative, more compact multiquark configurations governed by short-distance QCD dynamics. This possibility is reminiscent of the interplay between molecular and compact interpretations in the tetraquark sector, where different dynamical mechanisms may coexist or compete. Recent experimental progress in multi-body \(B\)-meson decays, such as the observation of rich structures in channels like \(B \to D^{(*)} D K\), further motivates the exploration of new configurations beyond conventional molecular descriptions.

In this work, we investigate a novel class of hexaquark configurations inspired by \(DDK\)-type systems, but conceptually distinct from three-body hadronic molecules. Specifically, we consider compact six-quark structures in which three color-octet quark--antiquark clusters couple to form an overall color-singlet state. Such hidden-charm configurations, although not directly observable as asymptotic states, can generate nontrivial short-distance correlations through nonperturbative gluon exchanges. The resulting ``color-octet $\times$ color-octet $\times$ color-octet'' configuration provides a new framework for exploring multiquark dynamics beyond the conventional meson--meson picture.

To study the possible existence of these states, we employ the method of QCD sum rules, a QCD-based nonperturbative approach that relates hadronic observables to quark and gluon degrees of freedom through the operator product expansion (OPE)~\cite{Shifman:1978bx,Reinders:1984sr,
Albuquerque:2013ija,
Govaerts:1984hc,
P.Col,Shifman:1978by}. We construct a set of independent local interpolating currents with the desired quantum numbers and evaluate the corresponding two-point correlation functions. By incorporating both perturbative contributions and nonperturbative condensates up to dimension ten, we analyze the resulting sum rules and extract information on the mass spectra of the proposed hexaquark configurations. Particular attention is paid to the convergence of the OPE, the stability of the Borel window, and the relative contribution of the pole term, which are essential for establishing the reliability of the analysis in systems with high operator dimensions.

Our study aims to provide quantitative theoretical predictions that can guide future experimental searches.
If such compact hexaquark states exist near the $D D K$ thresholds, they may couple strongly to nearby hadronic channels, leading to three-body $D D K$ final states or other channels with the same quantum numbers. These signatures can be probed through amplitude analyses of multi-body B-meson decays at the LHCb and Belle II experiments. More broadly, this work extends the application of QCD sum rules to a new class of multiquark systems and sheds light on the interplay between hadronic threshold effects and genuine multiquark correlations in QCD.

This paper is organized as follows. In Sec.~II, we present the formalism of QCD sum rules and construct the interpolating currents for the proposed hexaquark configurations. In Sec.~III, we perform the OPE calculations and numerical analyses to extract the mass spectrum. In Sec.~IV, we investigate the decay properties of the hexaquark states. Section~V is devoted to the conclusions.

\section{Formalism}
In this work, we employ the QCD sum rule method to systematically analyze the hexaquark state with quantum numbers $J^{P}=0^{-}$. In the QCD sum rule framework, the starting point is the construction of the two-point correlation function from an appropriate interpolating current. For the hexaquark state under consideration, this correlation function takes the form
\begin{eqnarray}
\Pi(q)=i\int d^{4}x\,e^{iq\cdot x}\langle 0|T\{j(x),j^{\dagger}(0)\}|0\rangle .
\end{eqnarray}

We construct hexaquark currents containing two charm quarks and four light quarks, whose color structure is given by the coupling of three color-octet clusters: $8_{[\bar{d}c]} \otimes 8_{[\bar{c}d]} \otimes 8_{[\bar{s}u]}$. The explicit forms are  
\begin{eqnarray}
\label{eq:2}  
j_1(x)&=&f^{ABC}\,\bigl[\bar{d}_{a}(x)\,i\gamma_{5}\,(\lambda^{A})_{ab}\,Q_{b}(x)\bigr]\,
                     \bigl[\bar{Q}_{c}(x)\,i\gamma_{5}\,(\lambda^{B})_{cd}\,d_{d}(x)\bigr]\,
                     \bigl[\bar{s}_{e}(x)\,i\gamma_{5}\,(\lambda^{C})_{ef}\,u_{f}(x)\bigr], \\[4pt]                   
j_2(x)&=&d^{ABC}\,\bigl[\bar{d}_{a}(x)\,i\gamma_{5}\,(\lambda^{A})_{ab}\,Q_{b}(x)\bigr]\,
                     \bigl[\bar{Q}_{c}(x)\,i\gamma_{5}\,(\lambda^{B})_{cd}\,d_{d}(x)\bigr]\,
                     \bigl[\bar{s}_{e}(x)\,i\gamma_{5}\,(\lambda^{C})_{ef}\,u_{f}(x)\bigr], \\[4pt]
j_3(x)&=&f^{ABC}\,\bigl[\bar{d}_{a}(x)\,\gamma_{\mu}\,(\lambda^{A})_{ab}\,Q_{b}(x)\bigr]\,
                     \bigl[\bar{Q}_{c}(x)\,\gamma^{\mu}\,(\lambda^{B})_{cd}\,d_{d}(x)\bigr]\,
                     \bigl[\bar{s}_{e}(x)\,i\gamma_{5}\,(\lambda^{C})_{ef}\,u_{f}(x)\bigr], \\[4pt]
j_4(x)&=&d^{ABC}\,\bigl[\bar{d}_{a}(x)\,\gamma_{\mu}\,(\lambda^{A})_{ab}\,Q_{b}(x)\bigr]\,
                     \bigl[\bar{Q}_{c}(x)\,\gamma^{\mu}\,(\lambda^{B})_{cd}\,d_{d}(x)\bigr]\,
                     \bigl[\bar{s}_{e}(x)\,i\gamma_{5}\,(\lambda^{C})_{ef}\,u_{f}(x)\bigr], \\[4pt]  
j_5(x)&=&f^{ABC}\,\bigl[\bar{d}_{a}(x)\,\gamma_{\mu}\,(\lambda^{A})_{ab}\,Q_{b}(x)\bigr]\,
                     \bigl[\bar{Q}_{c}(x)\,i\gamma_{5}\,(\lambda^{B})_{cd}\,d_{d}(x)\bigr]\,
                     \bigl[\bar{s}_{e}(x)\,\gamma^{\mu}\,(\lambda^{C})_{ef}\,u_{f}(x)\bigr], \\[4pt]
j_6(x)&=&d^{ABC}\,\bigl[\bar{d}_{a}(x)\,\gamma_{\mu}\,(\lambda^{A})_{ab}\,Q_{b}(x)\bigr]\,
                     \bigl[\bar{Q}_{c}(x)\,i\gamma_{5}\,(\lambda^{B})_{cd}\,d_{d}(x)\bigr]\,
                     \bigl[\bar{s}_{e}(x)\,\gamma^{\mu}\,(\lambda^{C})_{ef}\,u_{f}(x)\bigr],
\end{eqnarray}
where $a,b,c,d,e,f$ (running from $1$ to $3$) and $A,B,C$ (running from $1$ to $8$) are color indices, $\lambda^{A}$ are the Gell-Mann matrices, $Q$ denotes the charm quark field, $f^{ABC}$ is the totally antisymmetric structure constant of $SU(3)$, and $d^{ABC}$ is the corresponding symmetric structure constant.

In the framework of QCD sum rules, our theoretical formulation is based on the fundamental assumption of quark-hadron duality. This assumption states that the correlation function $\Pi(q^2)$, which describes hadronic properties, has two equivalent representations at appropriate energy scales: at the quark-gluon level, it can be systematically expressed via the operator product expansion (OPE) as the sum of perturbative contributions and non-perturbative vacuum condensate terms; alternatively, at the hadronic level, the same correlation function can be written in the form of a dispersion relation involving hadron masses, coupling constants, and continuum spectral integrals. 

By introducing the Borel transform, which suppresses continuum contributions and improves the convergence of the expansion, the two sides can be matched within a selected Borel window, enabling the extraction of physical quantities such as masses and coupling constants of hadronic states from first principles of QCD. This approach provides a powerful theoretical tool for studying hadronic structures in the non-perturbative QCD regime, particularly exotic hadrons such as multiquark states and molecular states. Thus, we have the following relation:
\begin{eqnarray}
\Pi^{\text{OPE}}(q^2)=\Pi^{\text{phen}}(q^2).
\end{eqnarray}

In calculating the spectral density on the OPE side, we introduce the full propagators of the heavy quark $Q$ in momentum space and the light quark $q$ in coordinate space, denoted as $S^Q_{ab}(p)$ and $S^q_{ab}(x)$, respectively,
\begin{eqnarray}
S^Q_{ab}(p)\!\!\!&=&\!\!\!\frac{i \delta_{ab}(p\!\!\!\slash + m_Q)}{p^2 - m_Q^2} - \frac{i}{4} \frac{g_s t^A_{ab} G^A_{\alpha\beta} }{(p^2 - m_Q^2)^2} [\sigma^{\alpha \beta}
(p\!\!\!\slash + m_Q)
+ (p\!\!\!\slash + m_Q) \sigma^{\alpha \beta}] \nonumber \\\!\!\!&+&\!\!\!  \frac{i\delta_{ab}m_Q  \langle g_s^2 G^2\rangle}{12(p^2 - m_Q^2)^3}\bigg[ 1 + \frac{m_Q (p\!\!\!\slash + m_Q)}{p^2 - m_Q^2} \bigg] \nonumber \\\!\!\!&+&\!\!\!\frac{i \delta_{ab}}{48} \bigg\{ \frac{(p\!\!\!\slash +
m_Q) [p\!\!\!\slash (p^2 - 3 m_Q^2) + 2 m_Q (2 p^2 - m_Q^2)] }{(p^2 - m_Q^2)^6}
\times (p\!\!\!\slash + m_Q)\bigg\} \langle g_s^3 G^3 \rangle \; ,\label{SQ}
\end{eqnarray}

\begin{eqnarray}
S_{ab}^{q}(x)\!\!\!&=&\!\!\!\frac{i \delta_{ab} x\!\!\!\slash }{2 \pi^{2} x^{4}}-\frac{m_{q} \delta_{ab}}{4 \pi^{2} x^{2}}-\frac{i g_s t^{A}_{ab} G^A_{\alpha\beta}}{32 \pi^{2} x^{2}} (\sigma_{\alpha \beta} x\!\!\!\slash +x\!\!\!\slash \sigma_{\alpha \beta})-\frac{\delta_{ab}}{12} \langle \bar{q} q \rangle+\frac{i \delta_{ab} x\!\!\!\slash}{48} m_{q}\langle \bar{q} q \rangle-\frac{\delta_{ab} x^{2}}{192}  \langle \bar{q} G q \rangle\nonumber\\&+&  \frac{i \delta_{ab} x^{2} x\!\!\!\slash}{1152} m_{q} \langle \bar{q} G q \rangle -\frac{t^{A}_{ab} \sigma_{\alpha \beta}}{192} \langle \bar{q} G q \rangle +
\frac{i t^{A}_{ab} }{768} (\sigma_{\alpha \beta} x\!\!\!\slash +x\!\!\!\slash \sigma_{\alpha \beta}) m_{q} \langle \bar{q} G q \rangle\;.\label{eq:10}
\end{eqnarray}
In these expressions, the vacuum condensate terms of the QCD operators are explicitly listed for both propagators. The complete propagators of the light and heavy quarks can be found in Refs.~\cite{Reinders:1984sr,Albuquerque:2013ija}. Note that the color index $A$ and the Lorentz indices $\alpha, \beta$ appearing in the last two terms of Eq.~\eqref{eq:10} are to be contracted with those of the gluon field $G^{A}_{\alpha\beta}$ radiated from another propagator. This contraction is required because that external gluon field joins the quark and antiquark located at these two terms to form a mixed condensate. For further details, see Ref.~\cite{Albuquerque:2013ija}.

\begin{figure}[!htbp]
  \centering
  \includegraphics[width=15cm]{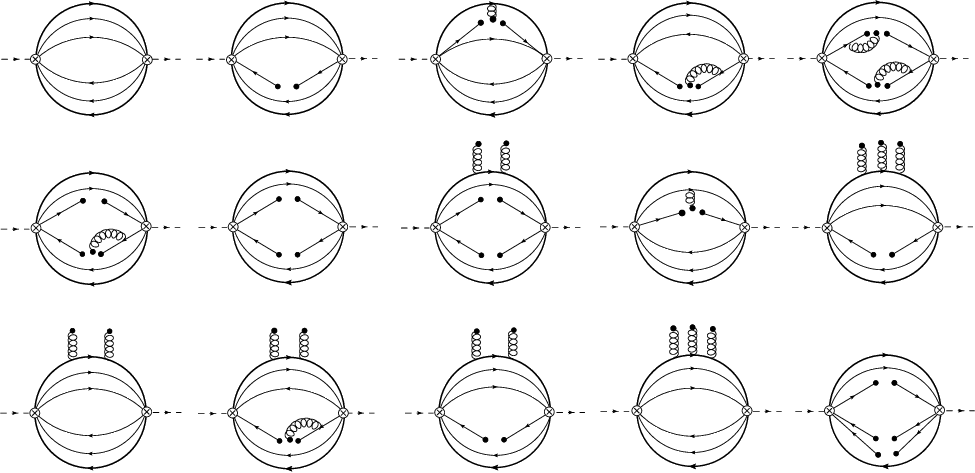}
  \vspace{1em}   % 图与图注的间距
  \setlength{\belowcaptionskip}{1em}   % 图注与正文的间距
  \caption{Shown here are the Feynman diagrams for the hexaquark state with three color-octet clusters, where only representative topologies are displayed and equivalent diagrams are omitted for clarity.}
  \label{Feyn-Diag}
\end{figure}

Using the full propagators, we obtain the spectral density $\rho^{\mathrm{OPE}}_i(s)$ up to condensates of dimension ten, and then the correlation function at the quark-gluon level can be determined via the dispersion relation
\begin{eqnarray}
  \Pi^{\text{OPE}}_{i} (q^2) &=& \int_{(2m_Q+m_s)^2}^{\infty} ds \frac{\rho_{i}^{\text{OPE}}(s)}{s - q^2},
\label{eq:11}
\end{eqnarray}
where $\rho^{\mathrm{OPE}}_i(s) = \operatorname{Im}[\Pi^{\mathrm{OPE}}_i(s)]/\pi$. In this and subsequent formulas, the subscript or superscript \(i\) takes values from 1 to 6, corresponding to the six different currents. For the hexaquark correlation function considered in this work, the representative leading-order diagrams contributing to the spectral density in Eq.~\eqref{eq:11} are displayed in Fig.~\ref{Feyn-Diag}.

On the phenomenological side, after isolating the ground-state contribution of the hexaquark state, the phenomenological representation of the correlation function $\Pi_i(q^2)$ can be written as the dispersion integral over the hadronic spectrum
\begin{eqnarray}
  \Pi_{i}^{\text{phen}}(q^2) = \frac{(\lambda^i_{X})^{2}} {(M_{X}^{i})^{2} - q^2 } +  \int_{s_0}^\infty \text{d}s \frac{\rho_{X}^{i}(s)}{s - q^2},
\end{eqnarray} 
where $M^i_X$ is the mass of the $J^P=0^-$ hexaquark state, $\rho^i_X(s)$ is its spectral density including contributions from higher excited states and the continuum, and $s_0$ is the threshold for excited states and the continuum. The coupling constant $\lambda_X$ is defined by $\langle 0|j_{X}^{i}|X\rangle = \lambda_{X}^{i}$, with $X$ denoting the lowest hexaquark state.

After applying the Borel transform to the phenomenological side of the correlation function, we obtain
\begin{eqnarray}
  \Pi_{i}^{\text{phen}}(s_0, M_B^2) = (\lambda^i_{X})^{2}e^{-(M_X^i)^2/M_B^2} + \int_{s_0}^\infty \text{d}s \rho^i_X(s) e^{-s/M_B^2}.
  \label{Pi-phen}
\label{eq:13}
\end{eqnarray}
Under the quark-hadron duality approximation, applying the Borel transform to
Eq.~\eqref{eq:11} and equating the result with Eq.~\eqref{eq:13} yields the master equation
\begin{eqnarray}
   (\lambda^{i}_{X})^{2}e^{-(M_X^i)^2/M_B^2}&=& \int_{(2m_Q+m_s)^2}^{s_0} \text{d}s \rho^{\text{OPE}}_i(s)e^{-s/M_B^2},
\end{eqnarray}
Next, we define the moments in the QCD sum rules
\begin{eqnarray}
  L_0(s_0, M_B^2) &=& \int_{(2m_{Q}+m_s)^2}^{s_0} \text{d}s \, \rho_i^{\text{OPE}}(s) e^{-s/M_B^2},
\label{eq:15}
\end{eqnarray}
\begin{eqnarray}
  L_1(s_0, M_B^2) &=& 
\frac{\partial }{\partial \frac{1}{M_B^2}}L_0(s_0, M_B^2).
\label{eq:16}
\end{eqnarray}
Finally, we can obtain the mass of the hexaquark state
\begin{eqnarray}
  M_{X}^{i}(s_{0},M_{B}^{2}) = \sqrt{-\frac{L_1^{\text{i}}(s_0, M_B^2)}{L_0^i(s_0, M_B^2)}}. \label{mass-equation}
  \label{eq:17}
\end{eqnarray}

\section{Numerical analysis}
We employ Eqs.~\eqref{eq:15}-\eqref{eq:17} for the numerical analysis of the hexaquark state with
$J^P = 0^-$. The following values for various QCD parameters are adopted, as suggested in Refs.~\cite{Reinders:1984sr, Shifman:1978bx, Shifman:1978by, Narison:1989aq, Colangelo:2000dp, ParticleDataGroup:2022pth} and listed in Table~\ref{QCDParam}.

\begin{table}[!htbp]
\centering
\setlength{\tabcolsep}{1.25pc}
\begin{tabular}{lc}
\hline
Parameter Name & Value \\
\hline
$m_u,m_d$ & $0$ \\
$m_c(m_c)=\overline{m_c}$ & $(1.27\pm0.02)~\text{GeV}$ \\
$m_s$ & $(0.13 \pm 0.03)~\text{GeV}$ \\
$\langle q \bar{q}\rangle$ & $-(0.24 \pm 0.01)^3~\text{GeV}^3$ \\
$\langle g_s^2 G^2\rangle$ & $0.88~\text{GeV}^4$ \\
$\langle g_s^3 G^3\rangle$ & $0.045~\text{GeV}^6$ \\
$\langle s \bar{s}\rangle / \langle q \bar{q}\rangle$ & $(0.8 \pm 0.1)$ \\
$m_0^2 \equiv \langle qG \bar{q}\rangle / \langle q \bar{q}\rangle$ & $(0.8\pm0.2)~\text{GeV}^2$ \\
\hline
\end{tabular}
\caption{QCD input parameters used in the numerical calculations of this paper~\cite{Reinders:1984sr, Shifman:1978bx, Shifman:1978by, Narison:1989aq, Colangelo:2000dp, ParticleDataGroup:2022pth}.}
\label{QCDParam}
\end{table}

In the framework of QCD sum rules, the Borel parameter \(M_B^2\) is introduced to suppress contributions from higher excited states and the continuum while simultaneously enhancing the convergence of the operator product expansion (OPE). In principle, if the OPE were fully convergent and the continuum model were perfectly accurate, the hadron mass \(M_X\) extracted from the sum rules would be independent of \(M_B^2\). However, in practical applications, truncations of the OPE at finite-dimensional operators and the phenomenological parameterization of the continuum threshold \(s_0\) introduce a residual dependence of \(M_X\) on \(M_B^2\). 

To ensure the reliability of results extracted from sum rules, it is essential to identify an appropriate Borel window \(M_B^2 \in [(M_B^{2})_\text{min}, (M_B^{2})_\text{max}]\), within which physical quantities exhibit minimal sensitivity to \(M_B^2\), the OPE demonstrates good convergence, and the continuum contribution remains adequately suppressed. The determination of this Borel window is guided by two complementary criteria~\cite{Reinders:1984sr, Shifman:1978bx, Shifman:1978by, Colangelo:2000dp}.

The first criterion is the OPE convergence requirement, which stipulates that contributions from higher-dimensional operators should be substantially smaller than the total OPE contribution to ensure the validity of the truncation. This condition can be expressed quantitatively as
\begin{equation}
  R_i^{\text{cond}}(s_0, M_B^2) = \frac{L_i^{\text{dim}}(s_0, M_B^2)}{L_0(s_0, M_B^2)} ,
\end{equation}
where \(L_0(s_0, M_B^2)\) represents the total contribution to the Borel-transformed sum rules, and \(L_i^{\text{dim}}(s_0, M_B^2)\) denotes the contribution from the \(i\)-th higher-dimensional operator in the OPE. Quantitatively, we require that the dimension-ten contribution satisfies
\begin{equation}
  \left| R_{i}^{\text{dim}=10}(s_0,M_B^2) \right| \leq 10\% .
\end{equation}

The second criterion is the pole contribution, which requires that the contribution associated with the lowest-lying hadronic state remains sufficiently large in the Borel-transformed correlation function after the continuum subtraction. In this way, the contamination from higher excited states and the continuum can be kept under control. Quantitatively, the pole contribution (PC) is defined as
\begin{equation}
  R^{\text{PC}}(s_0,M_B^2)
  =
  \frac{L_0(s_0,M_B^2)}{L_0(\infty,M_B^2)} \, ,
  \label{RatioPC}
\end{equation}
where \(L_0(s_0,M_B^2)\) denotes the Borel moment with the continuum contribution above \(s_0\) subtracted, while \(L_0(\infty,M_B^2)\) represents the corresponding moment without continuum subtraction. Since the continuum contribution generally increases with increasing \(M_B^2\), this requirement provides an upper bound \((M_B^2)_{\rm max}\) for the Borel window.

In practical analyses of multiquark systems, and in particular hexaquark states, the pole contribution is often smaller than that in conventional meson or baryon sum rules. Following the widely accepted empirical criterion~\cite{Zhang:2025qmg} for isolating ground-state dominance in hexaquark studies, we require
\begin{equation}
  R^{\text{PC}}(s_0,M_B^2) \geq 15\% ,
\end{equation}
which ensures that the extracted mass is still governed by a sufficiently sizable ground-state contribution within the chosen Borel window.

In practical calculations, the determination of the continuum threshold parameter \(\sqrt{s_0}\) generally follows two widely adopted criteria, as outlined in Refs.~\cite{Finazzo:2011he, Qiao:2013raa, Qiao:2013dda}. First, the Borel mass curve \(M_X(M_B^2)\) must exhibit a well-defined plateau within the chosen Borel window, indicating stability and insensitivity to variations in the Borel parameter \(M_B^2\). Second, the difference between the continuum threshold energy \(\sqrt{s_0}\) and the extracted mass \(M_X\) should fall within the range \(\sqrt{s_0} - M_X \in [0.4,\,0.8]\,\text{GeV}\), consistent with physical expectations.

To determine the optimal value of \(\sqrt{s_0}\), a point-by-point scan is performed over candidate intervals for \(\sqrt{s_0}\). Among these, parameter points that simultaneously satisfy both criteria are considered. The value of \(\sqrt{s_0}\) corresponding to the minimal fluctuation in the plateau is selected as the central value. After fixing the central value, the two boundary values obtained by shifting \(\sqrt{s_0}\) by \(\pm 0.2\,\text{GeV}\) are adopted to estimate the uncertainty.

\begin{figure}[!htbp]
  \centering
  \includegraphics[width=0.5\textwidth]{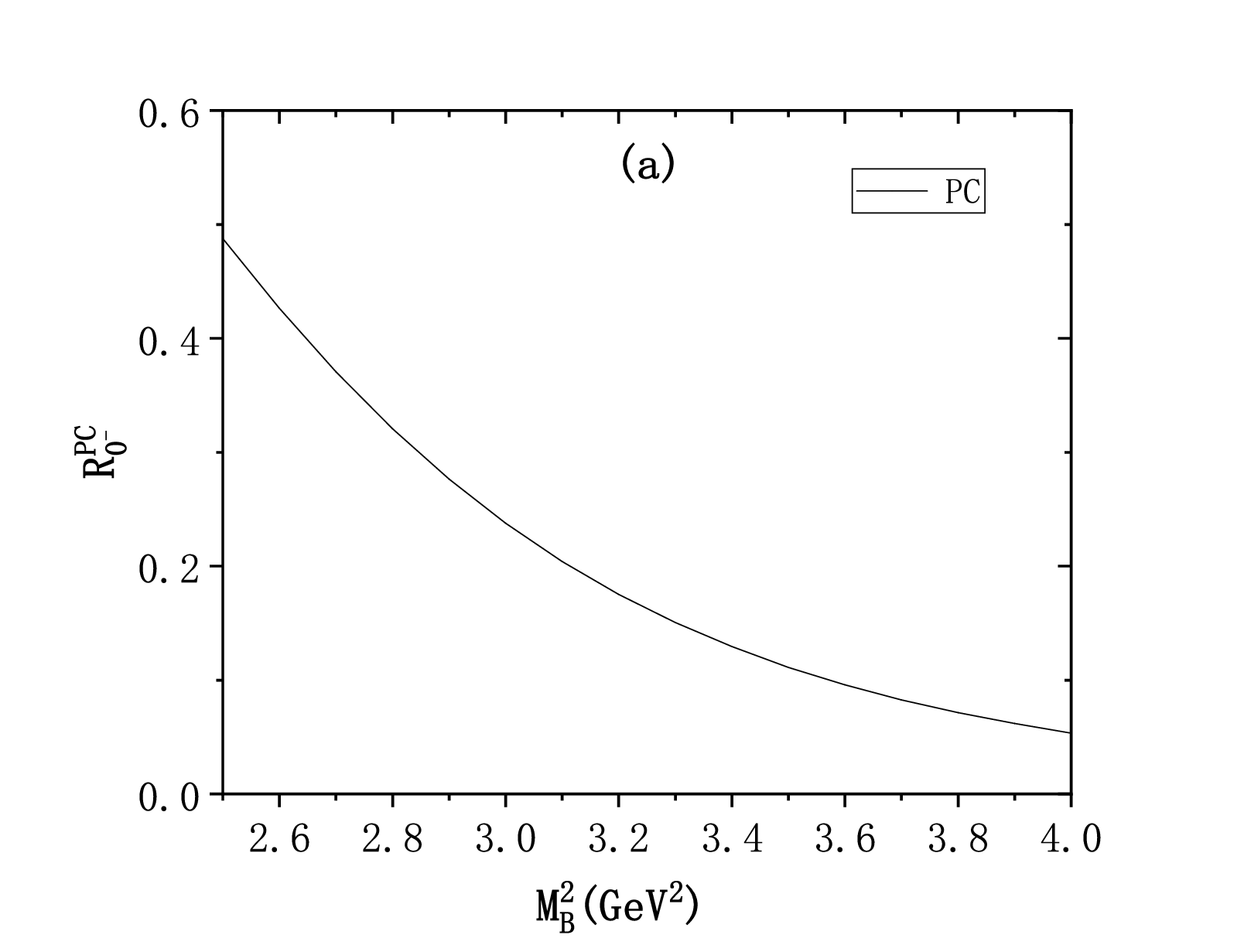}\hfill
  \includegraphics[width=0.5\textwidth]{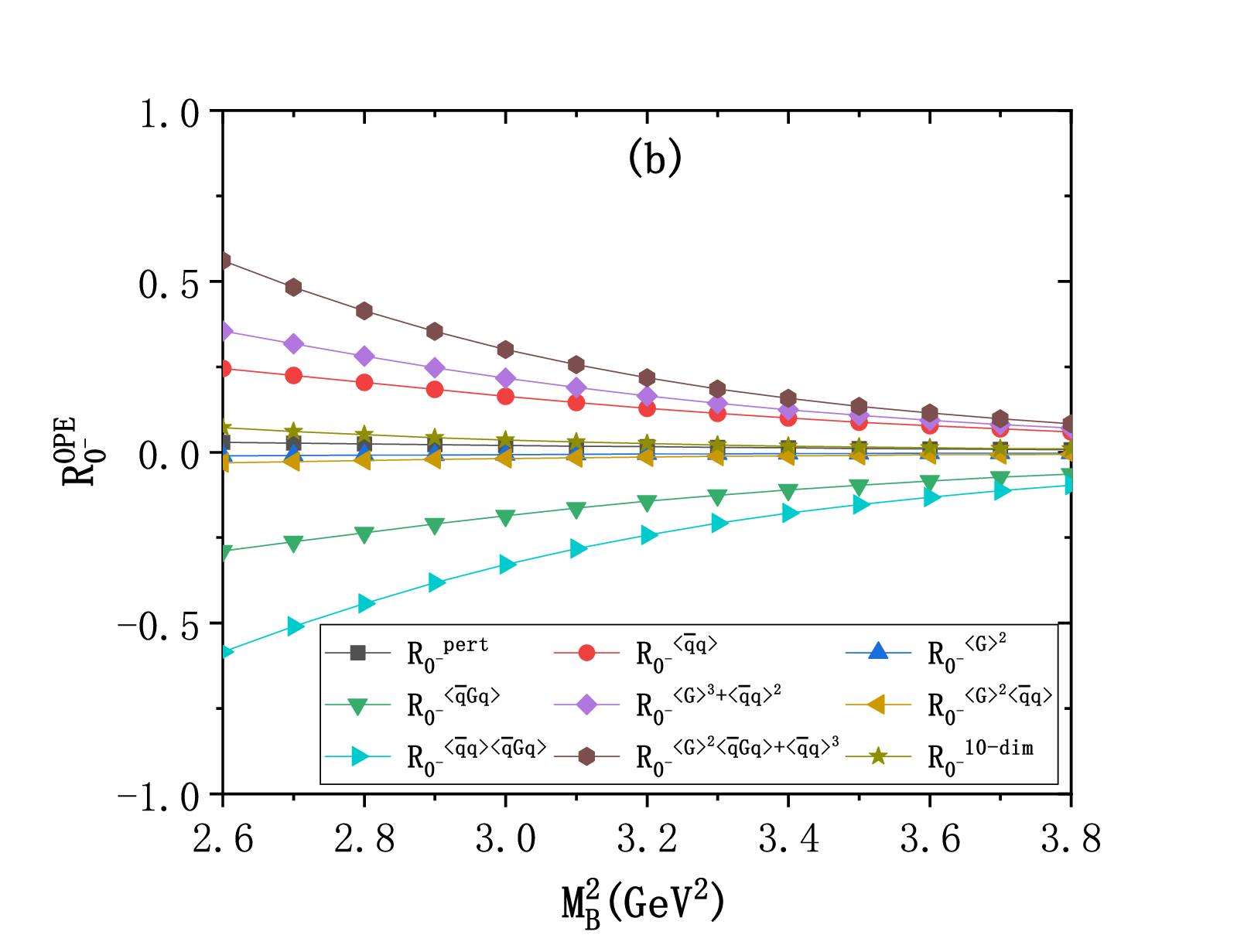}
  \caption{The figures for current $j_1$. (a) The pole contribution ratio $R_{0^-}^{\text{PC}}$ as a function of the Borel parameter $M_B^2$ with the central value of $s_0$; (b) OPE contribution ratio $R_{0^-}^{\text{OPE}}$ as functions of the Borel parameter $M_B^2$ with the central value of $s_0$.}
  \label{Feyn1}
\end{figure}

\begin{figure}[!htbp]
  \centering
  \includegraphics[width=0.5\textwidth]{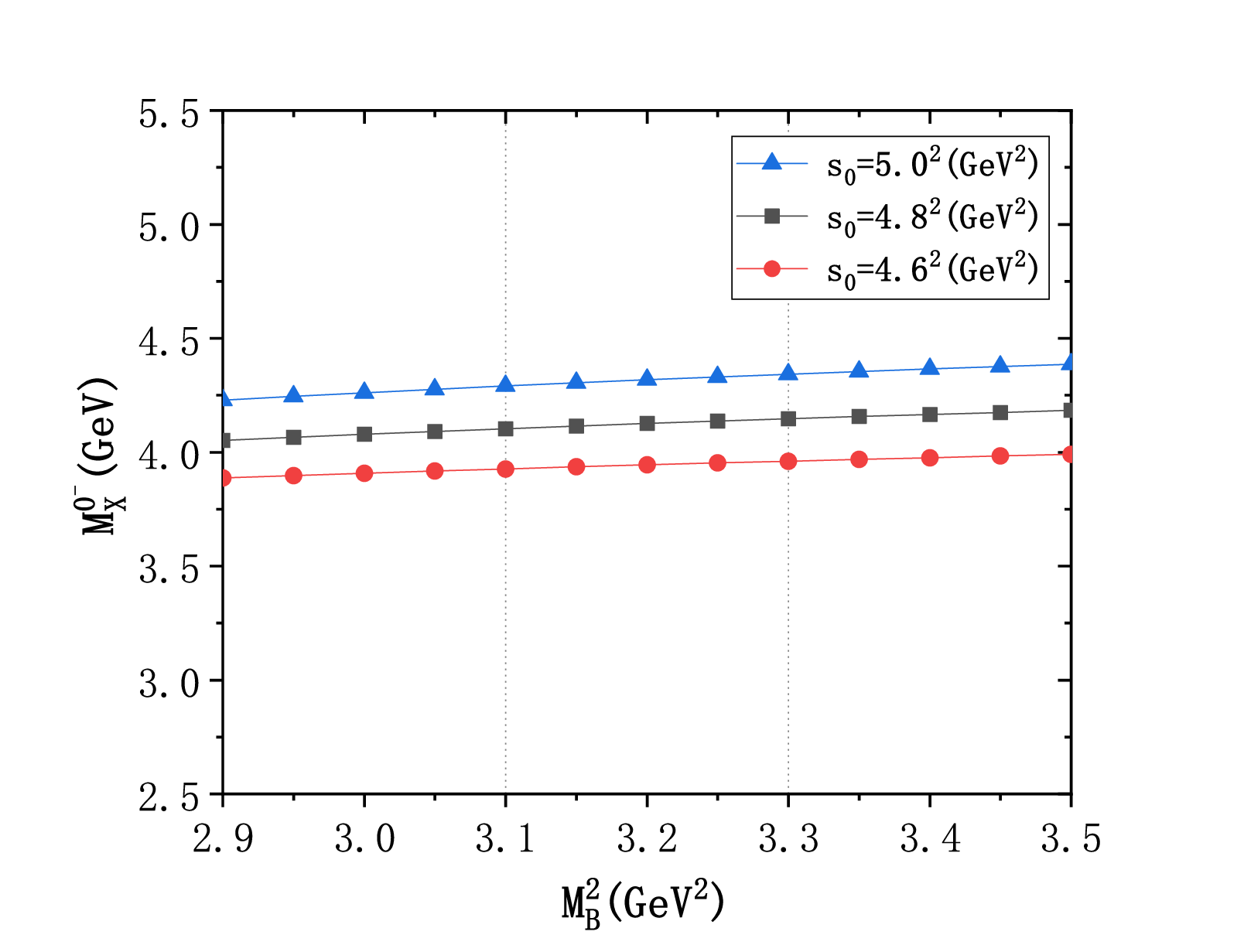}
  \caption{The masses $M_X^{0^-}$ as functions of $M_B^2$ for $s_0 = 4.60^2 \, \text{GeV}^2$, $4.80^2 \, \text{GeV}^2$, and $5.00^2 \, \text{GeV}^2$ from down to up, respectively, and the two vertical lines indicate the upper and lower bounds of valid Borel window with the central value of $s_0$.}
  \label{Feyn2}
\end{figure}

\begin{figure}[!htbp]
  \centering
  \includegraphics[width=0.5\textwidth]{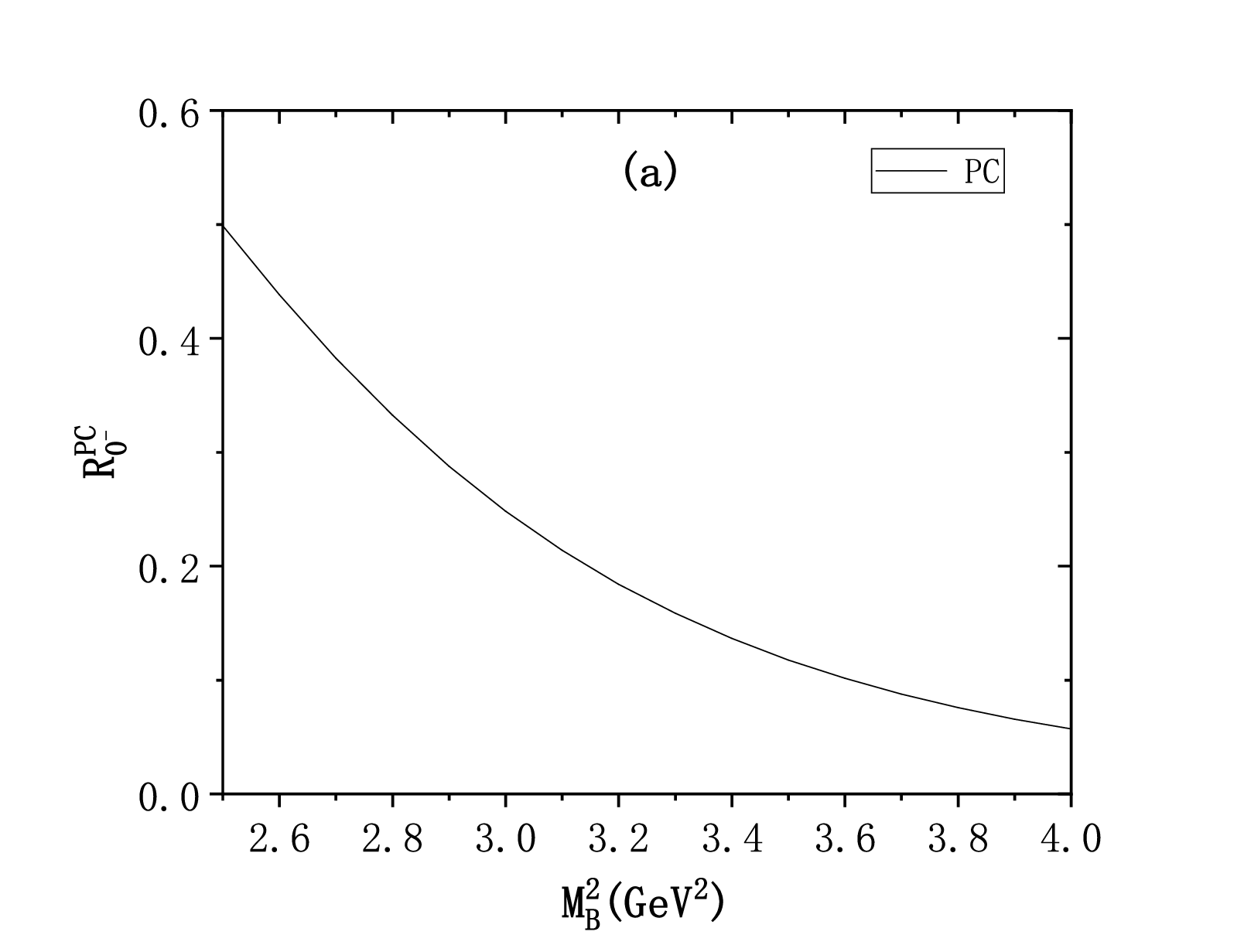}\hfill
  \includegraphics[width=0.5\textwidth]{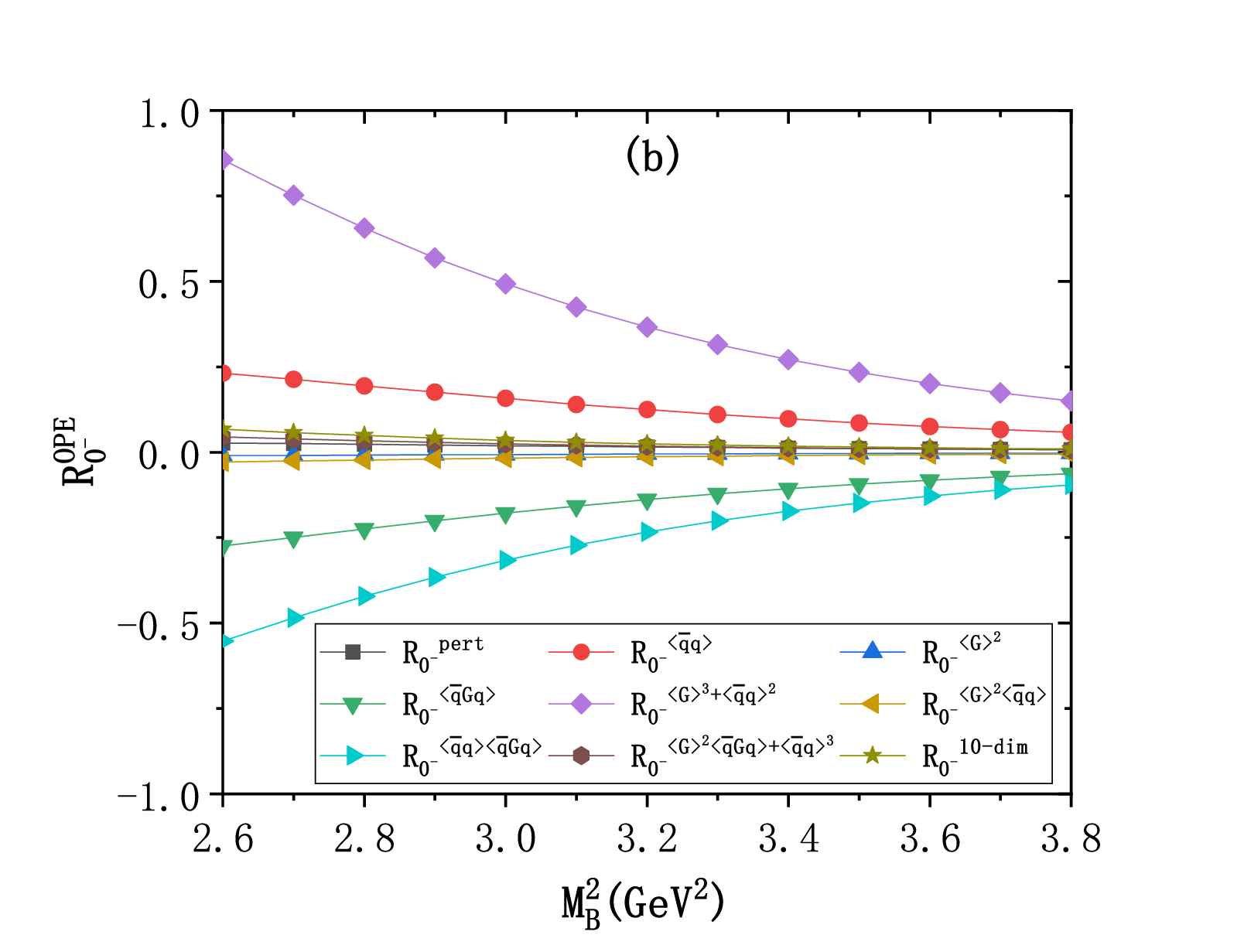}
  \caption{The same caption as in Fig.~\ref{Feyn1}, but for current $j_{2}$.}
  \label{Feyn3}
\end{figure}

\begin{figure}[!htbp]
  \centering
  \includegraphics[width=0.5\textwidth]{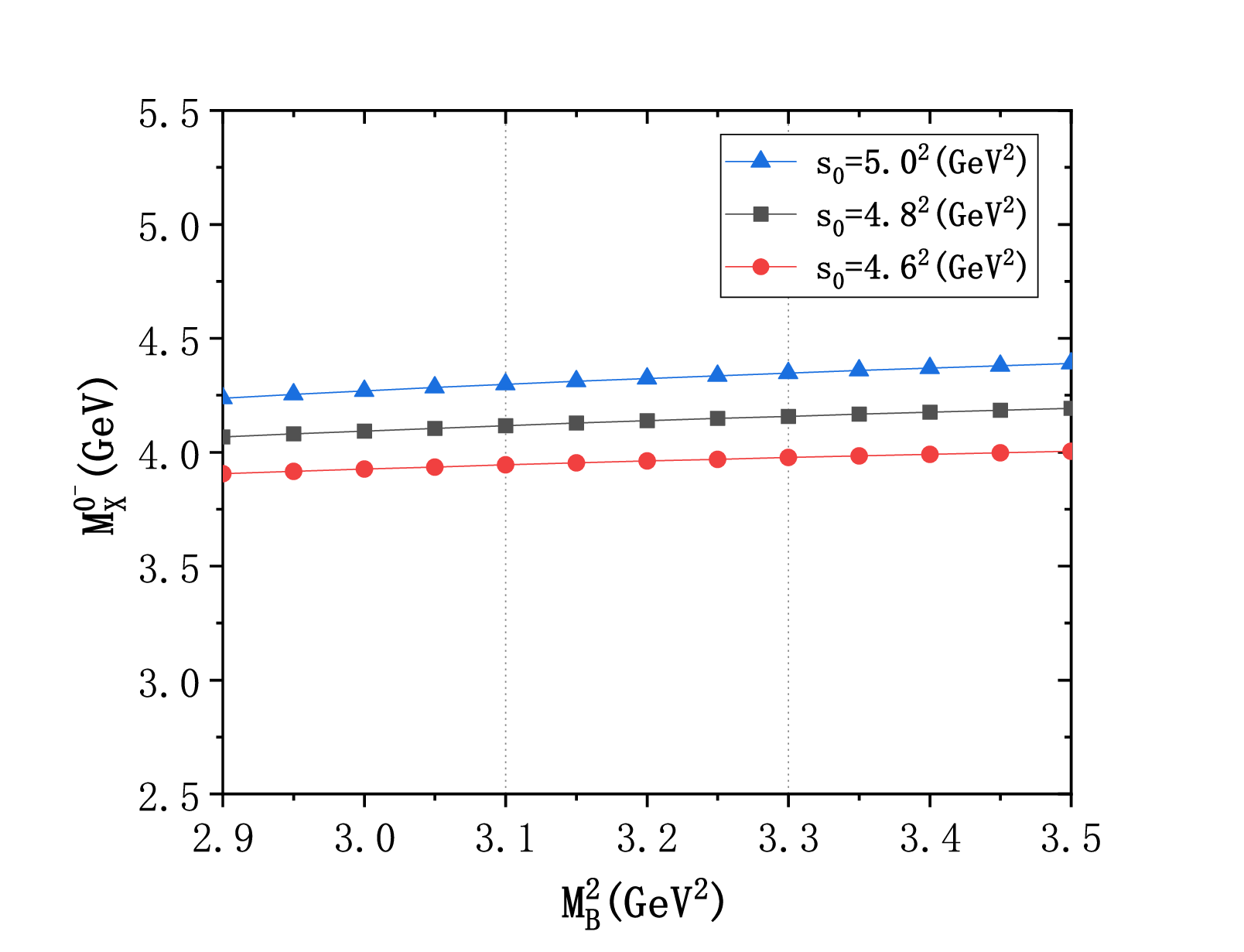}
  \caption{The same caption as in Fig.~\ref{Feyn2}, but for current $j_{2}$.}
  \label{Feyn4}
\end{figure}

\begin{figure}[!htbp]
  \centering
  \includegraphics[width=0.5\textwidth]{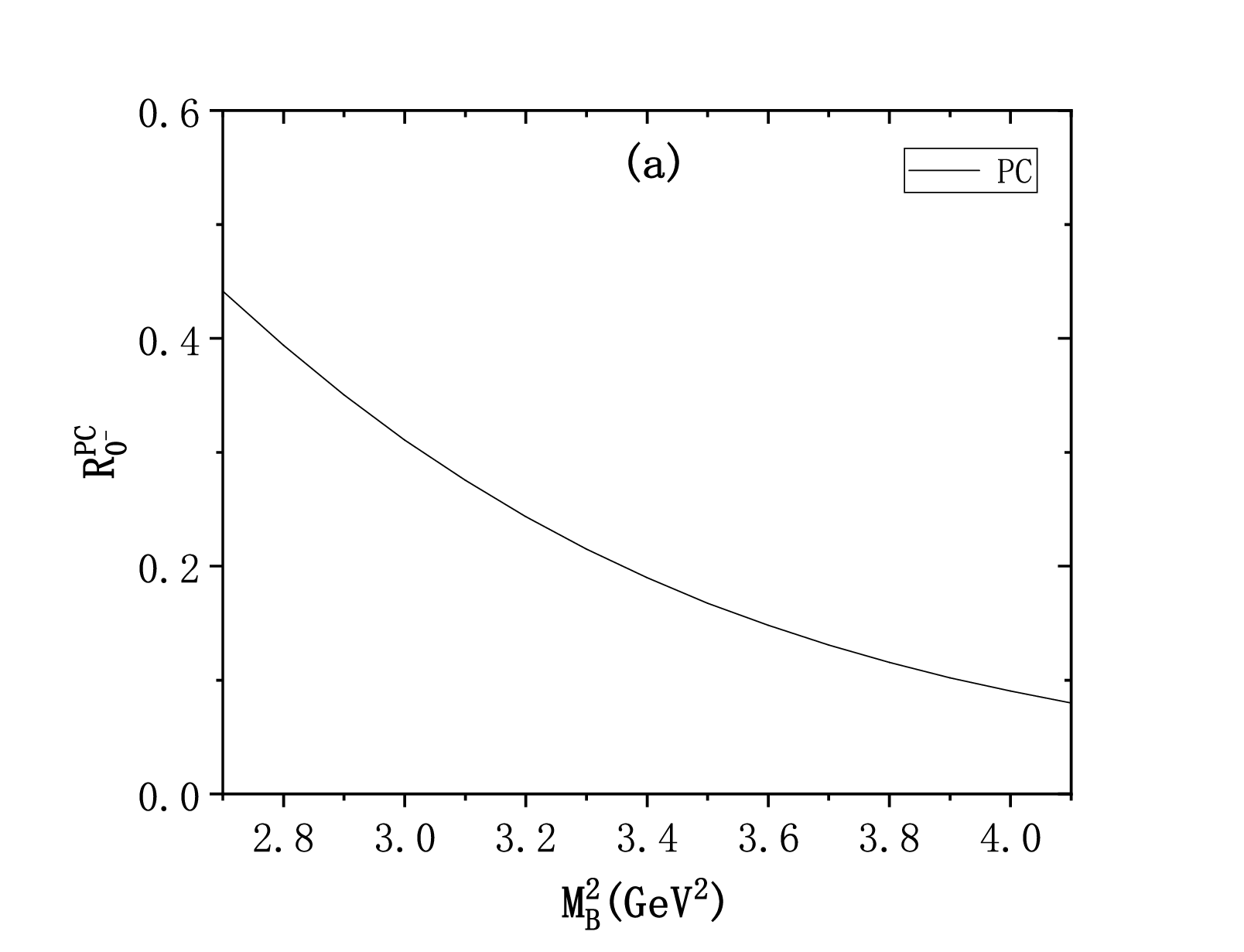}\hfill
  \includegraphics[width=0.5\textwidth]{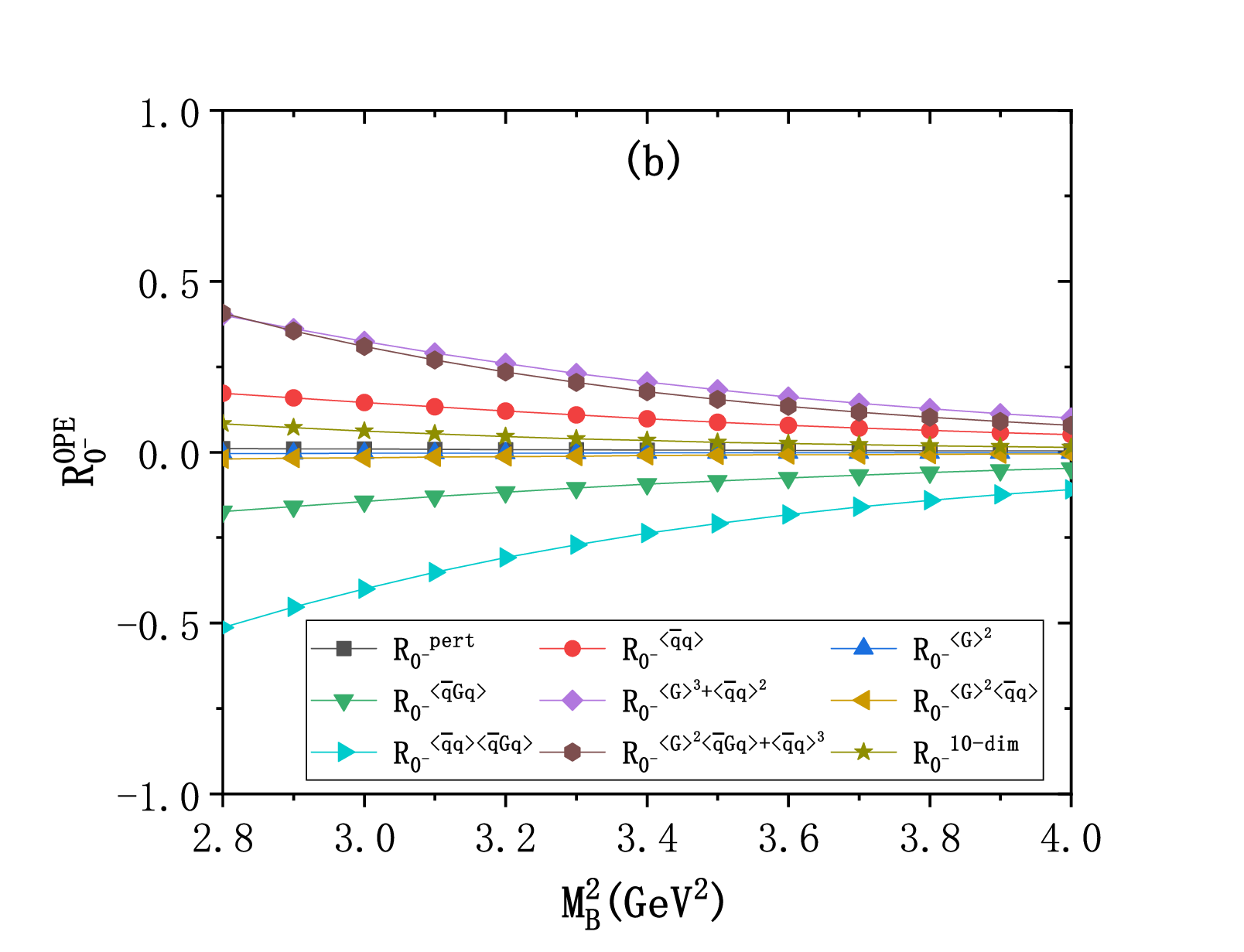}
  \caption{The same caption as in Fig.~\ref{Feyn1}, but for current $j_{3}$.}
  \label{Feyn5}
\end{figure}

\begin{figure}[!htbp]
  \centering
  \includegraphics[width=0.5\textwidth]{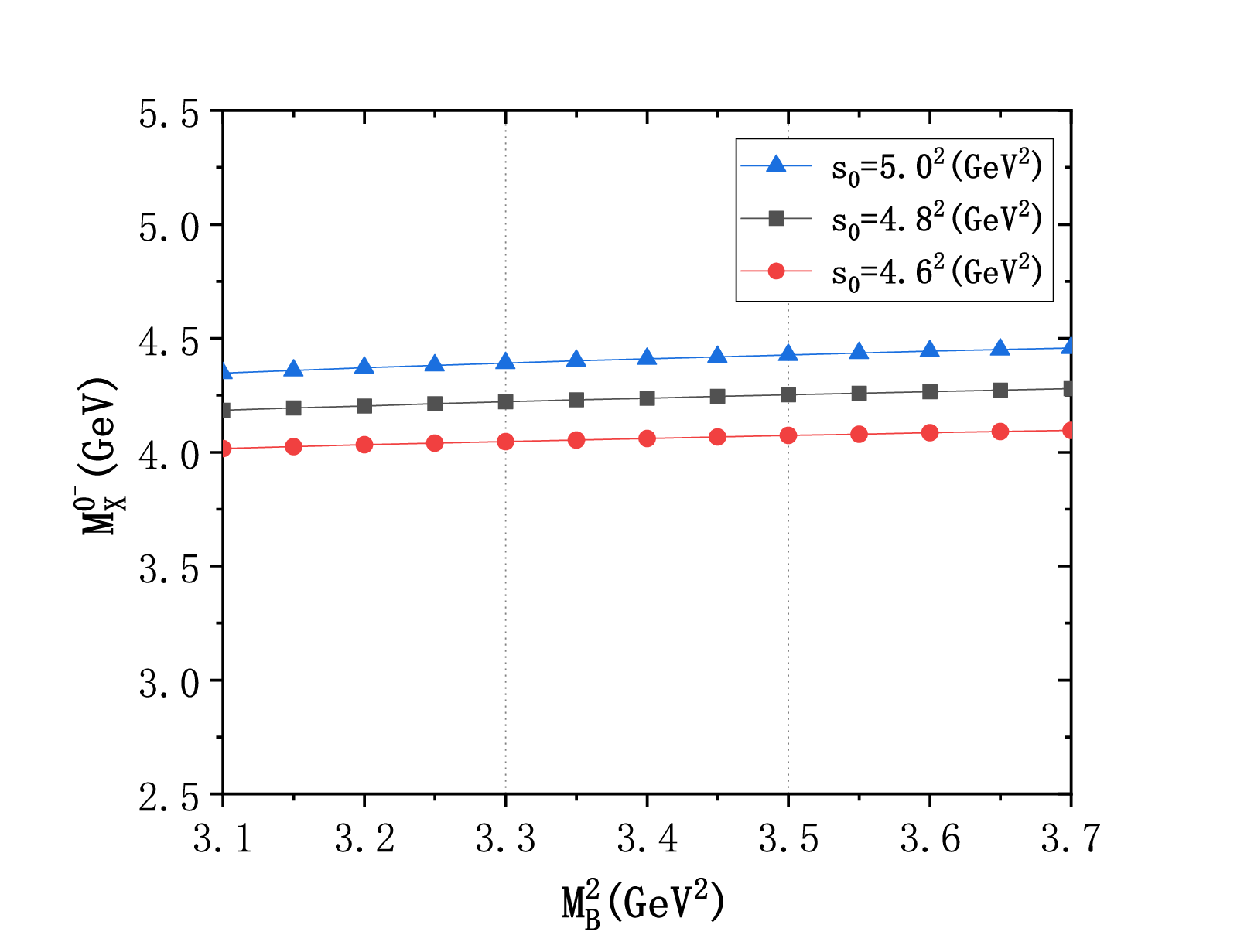}
  \caption{The same caption as in Fig.~\ref{Feyn2}, but for current $j_{3}$.}
  \label{Feyn6}
\end{figure}

\begin{figure}[!htbp]
  \centering
  \includegraphics[width=0.5\textwidth]{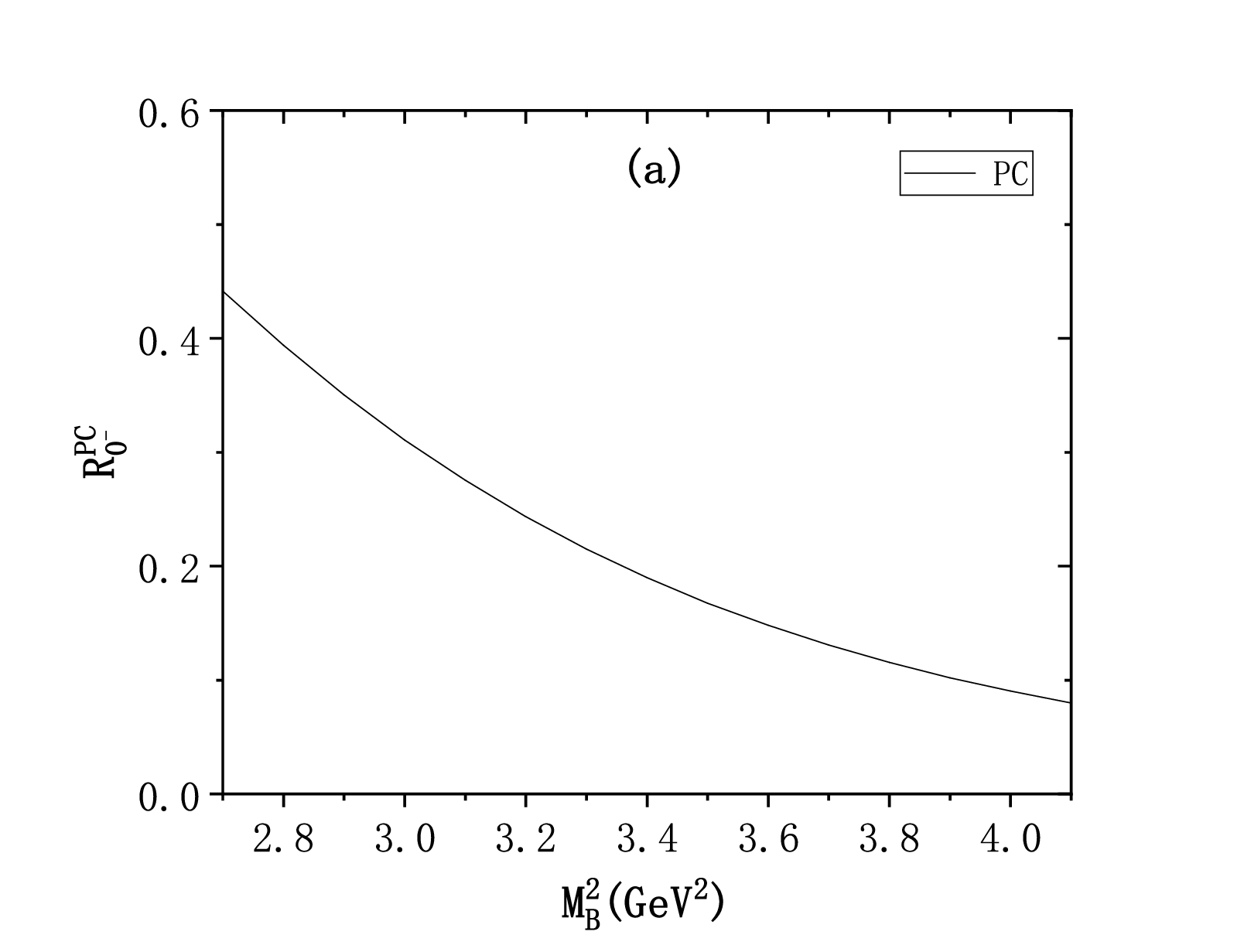}\hfill
  \includegraphics[width=0.5\textwidth]{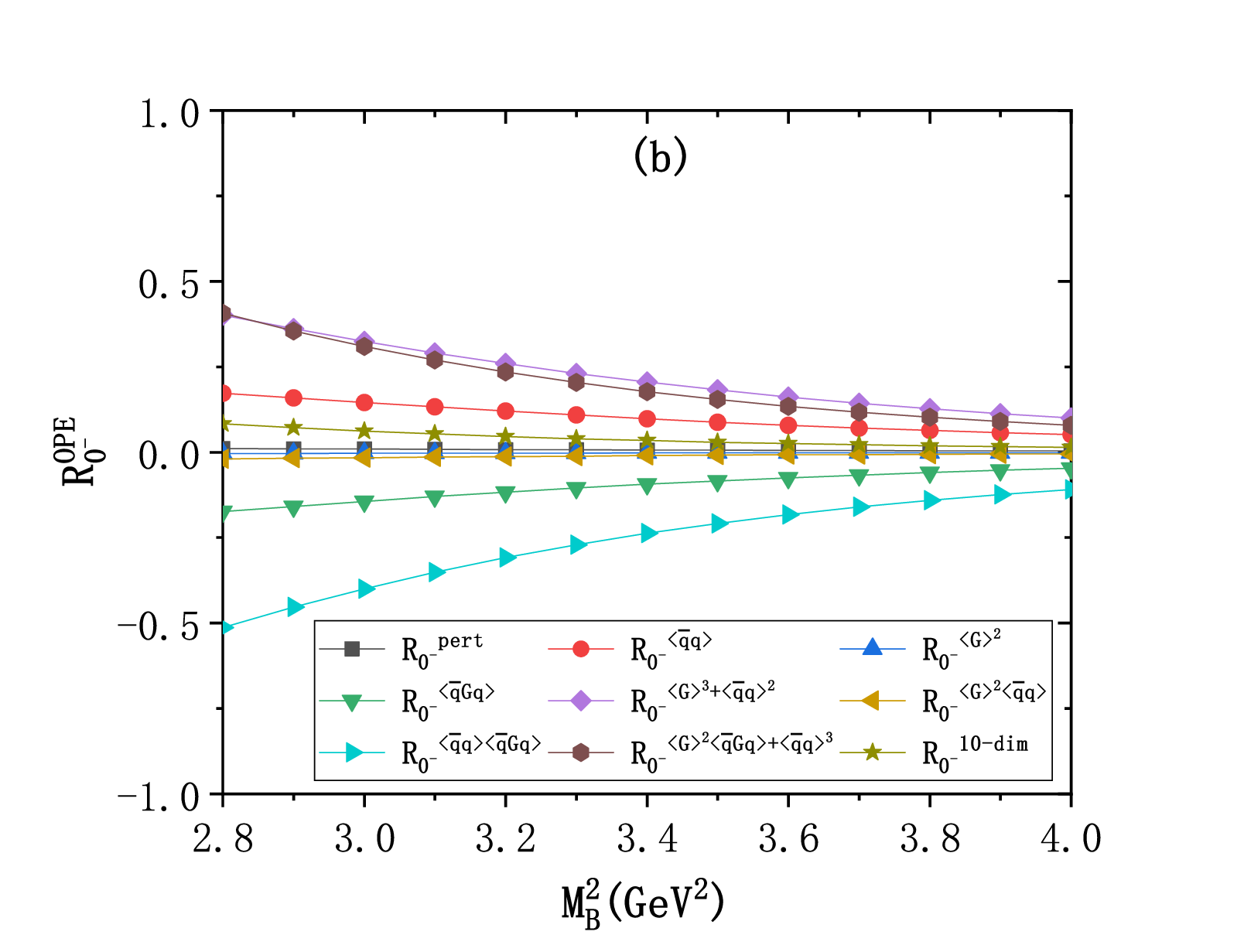}
  \caption{The same caption as in Fig.~\ref{Feyn1}, but for current $j_{4}$.}
  \label{Feyn7}
\end{figure}

\begin{figure}[!htbp]
  \centering
  \includegraphics[width=0.5\textwidth]{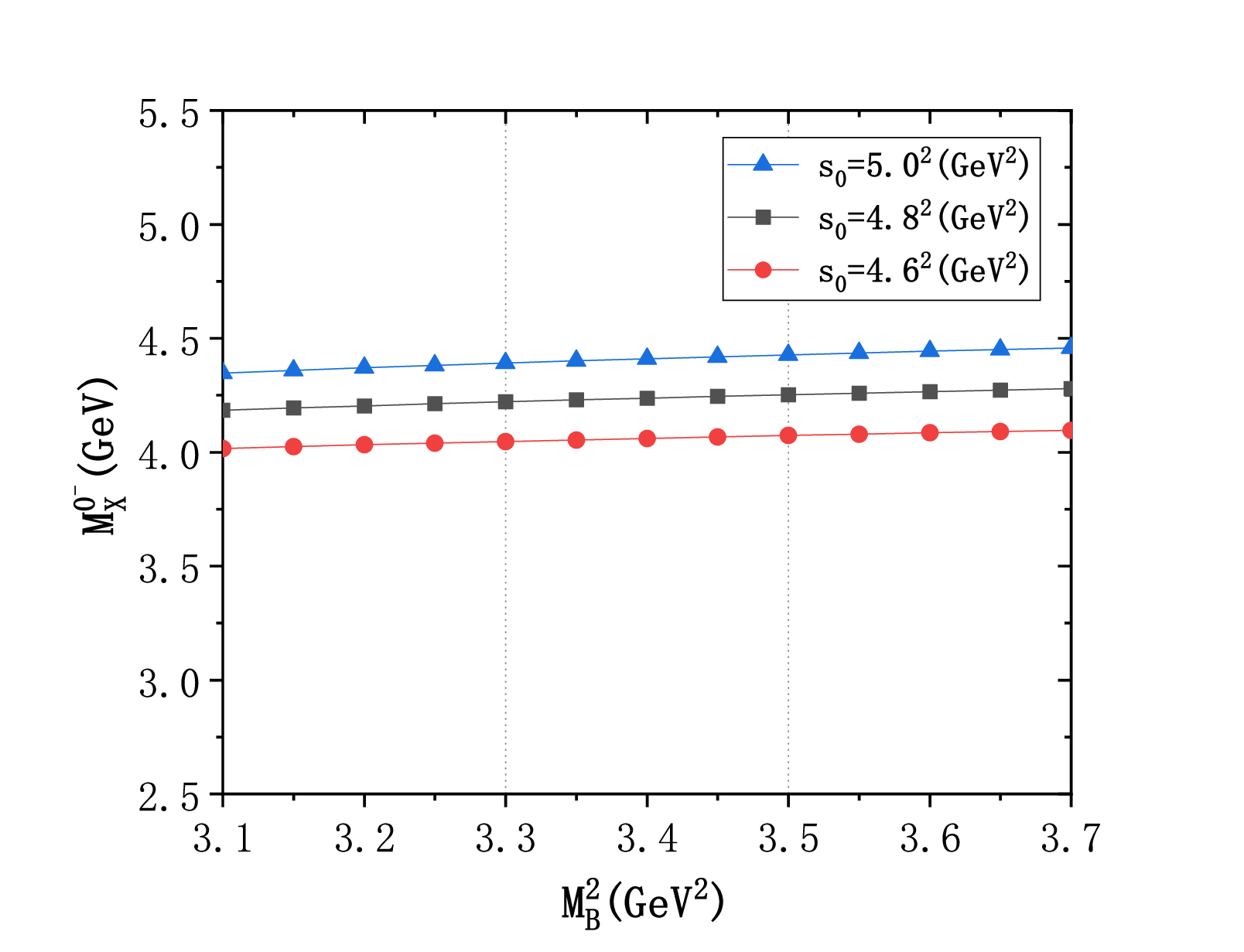}
  \caption{The same caption as in Fig.~\ref{Feyn2}, but for current $j_{4}$.}
  \label{Feyn8}
\end{figure}

\begin{figure}[!htbp]
  \centering
  \includegraphics[width=0.5\textwidth]{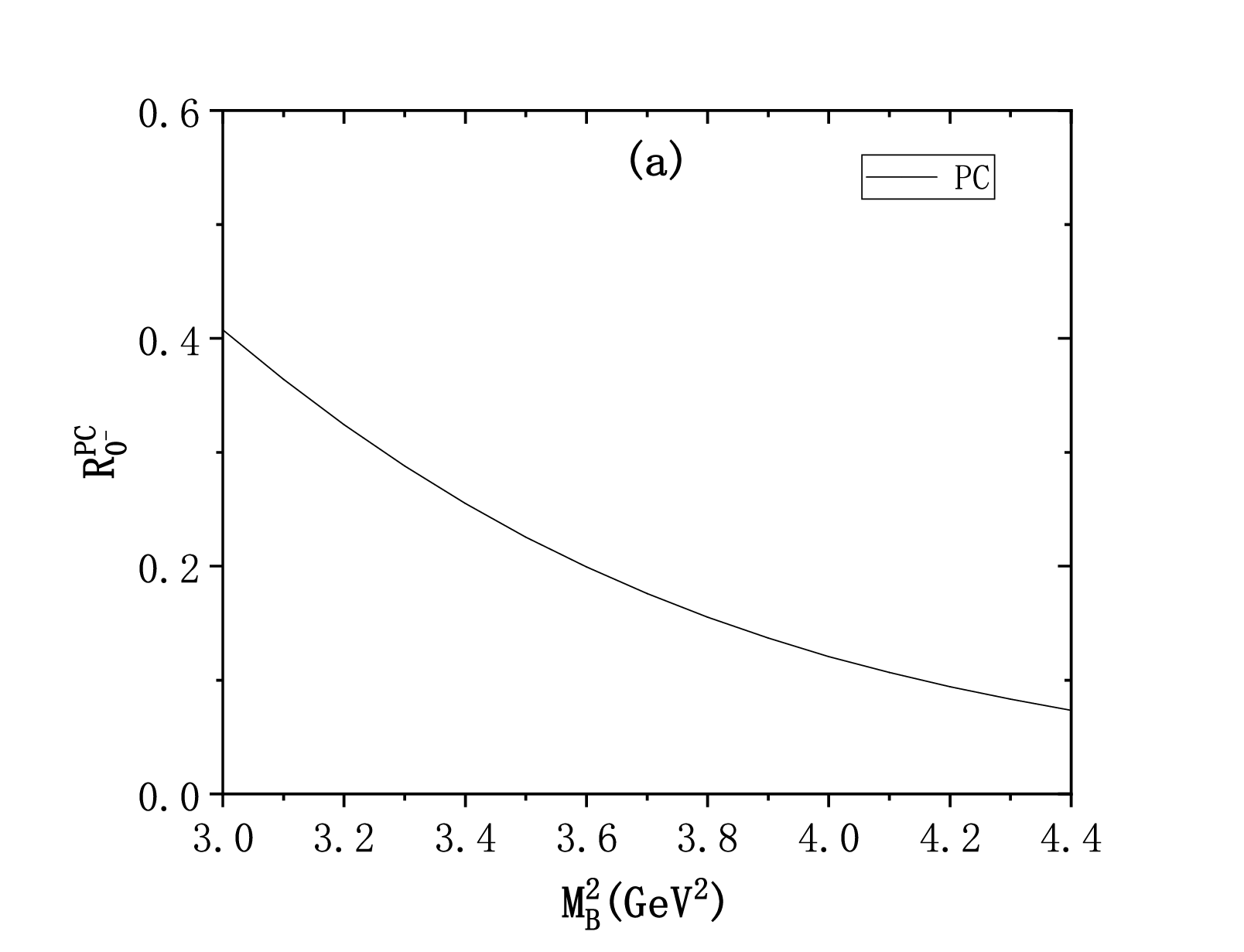}\hfill
  \includegraphics[width=0.5\textwidth]{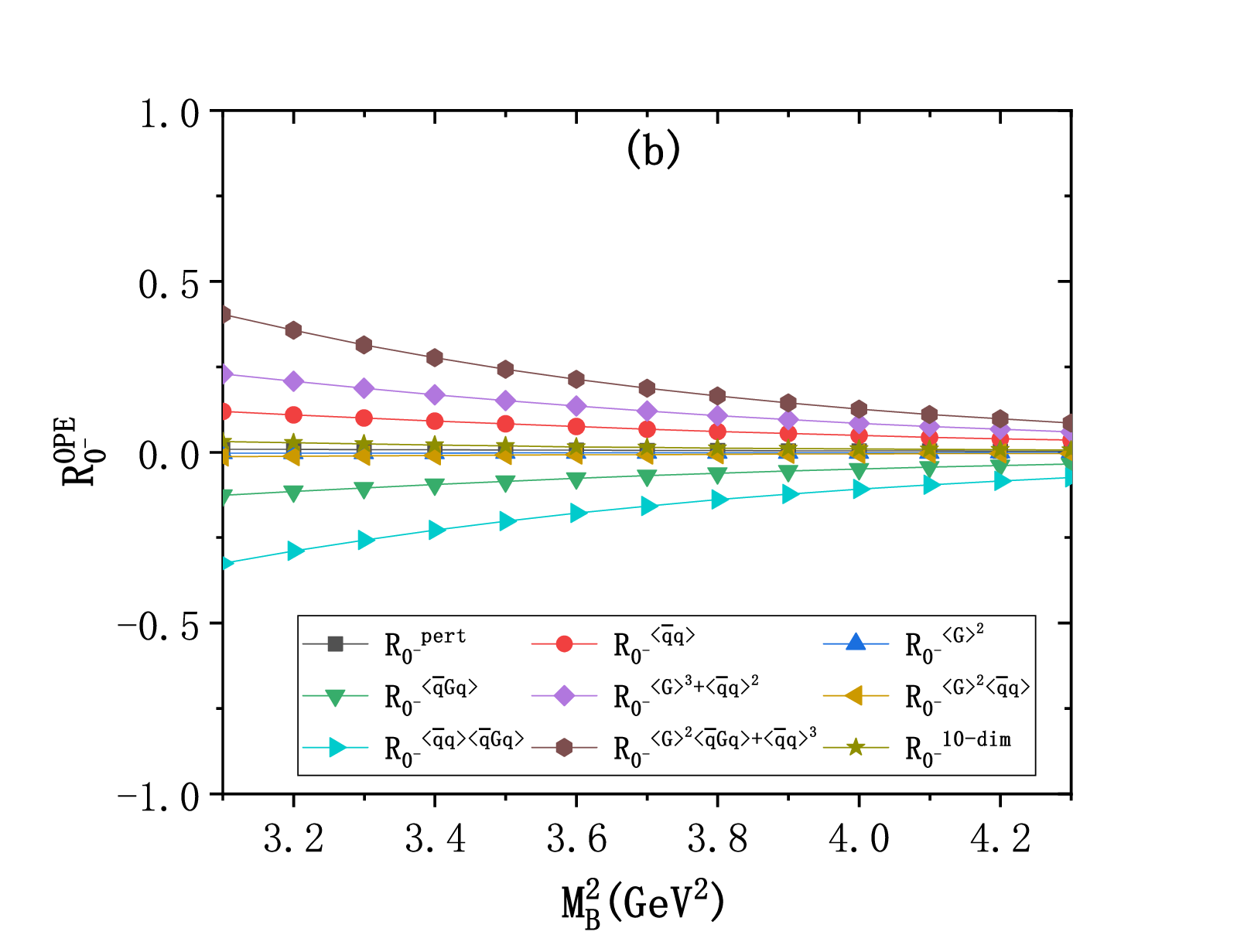}
  \caption{The same caption as in Fig.~\ref{Feyn1}, but for current $j_{5}$.}
  \label{Feyn9}
\end{figure}

\begin{figure}[!htbp]
  \centering
  \includegraphics[width=0.5\textwidth]{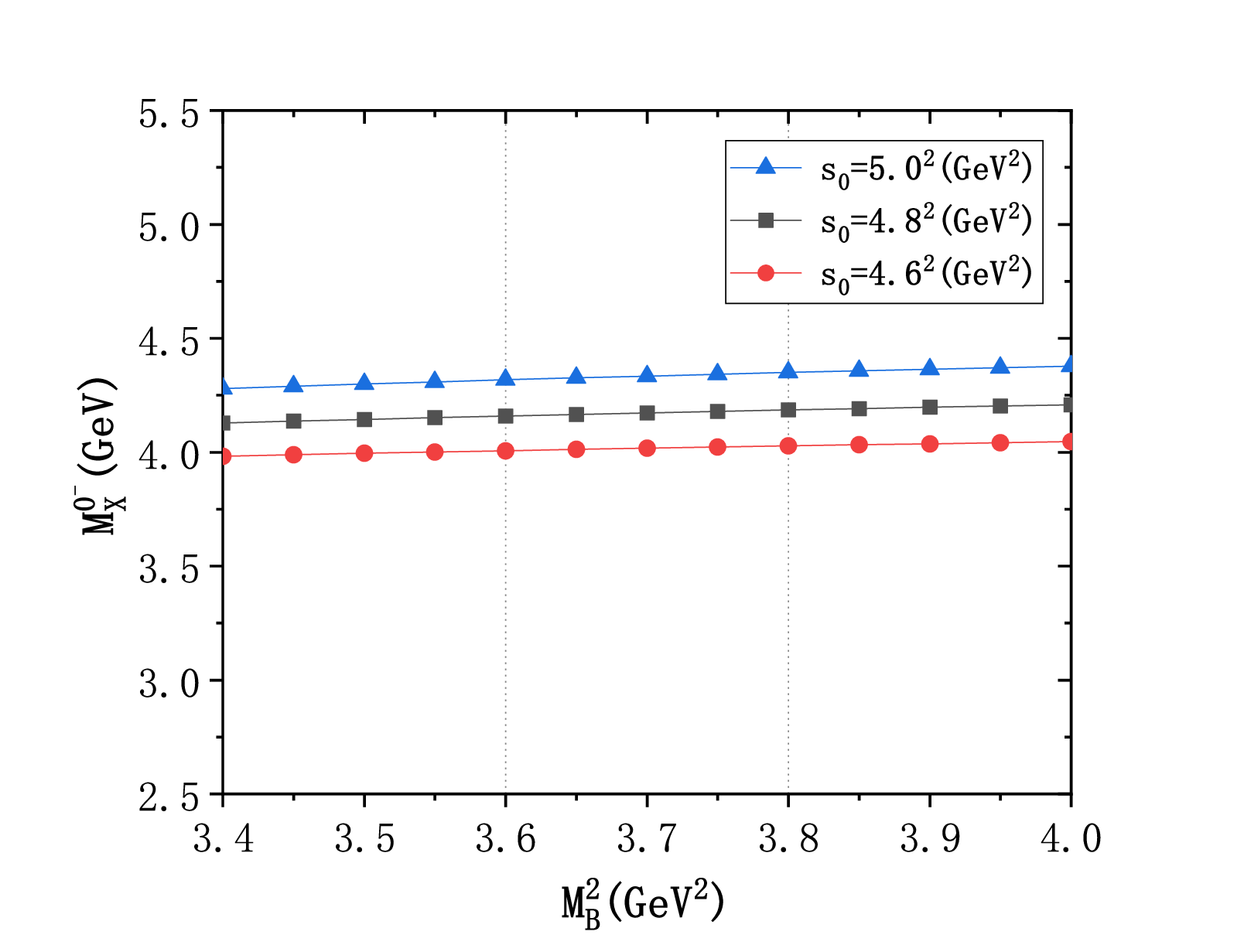}
  \caption{The same caption as in Fig.~\ref{Feyn2}, but for current $j_{5}$.}
  \label{Feyn10}
\end{figure}

\begin{figure}[!htbp]
  \centering
  \includegraphics[width=0.5\textwidth]{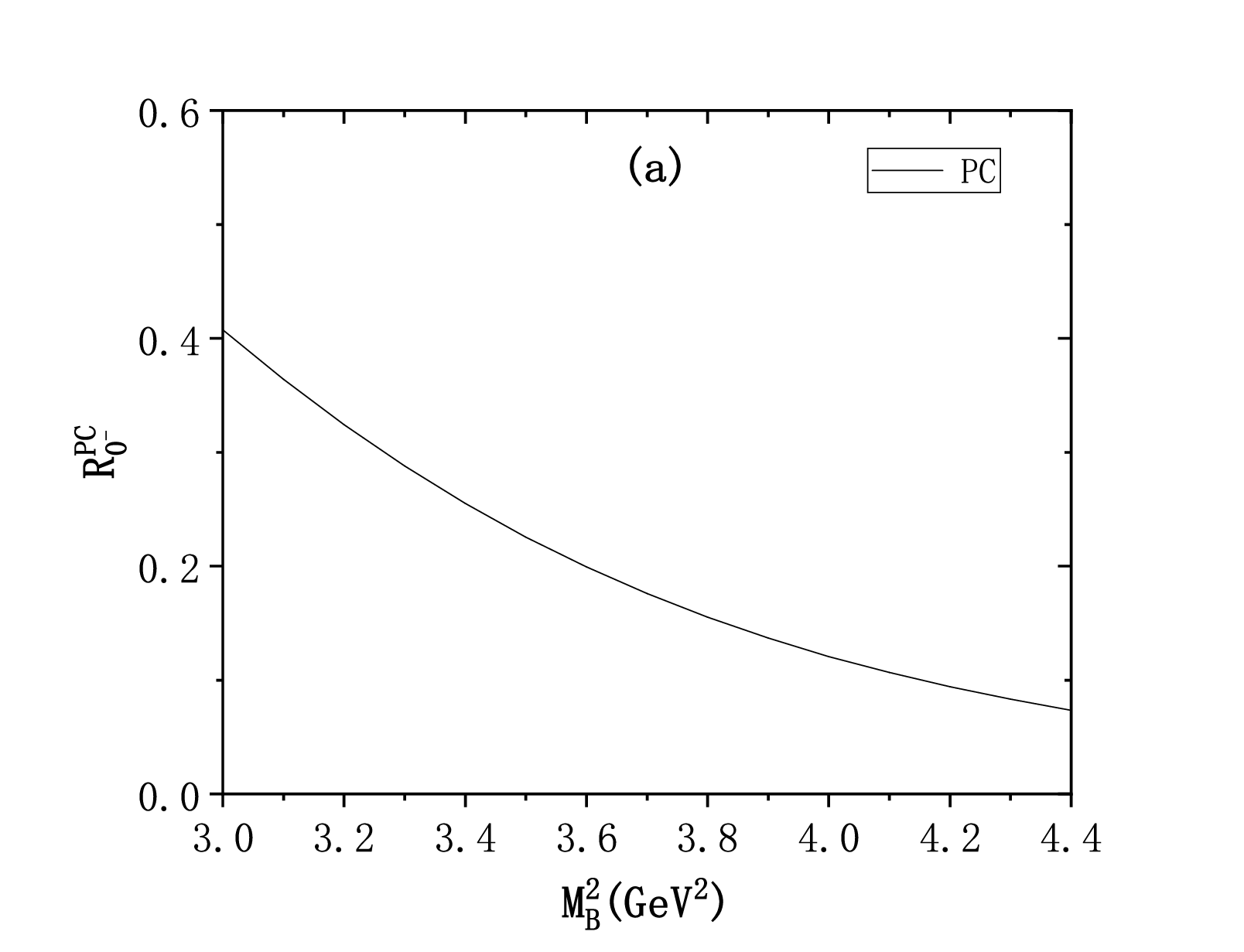}\hfill
  \includegraphics[width=0.5\textwidth]{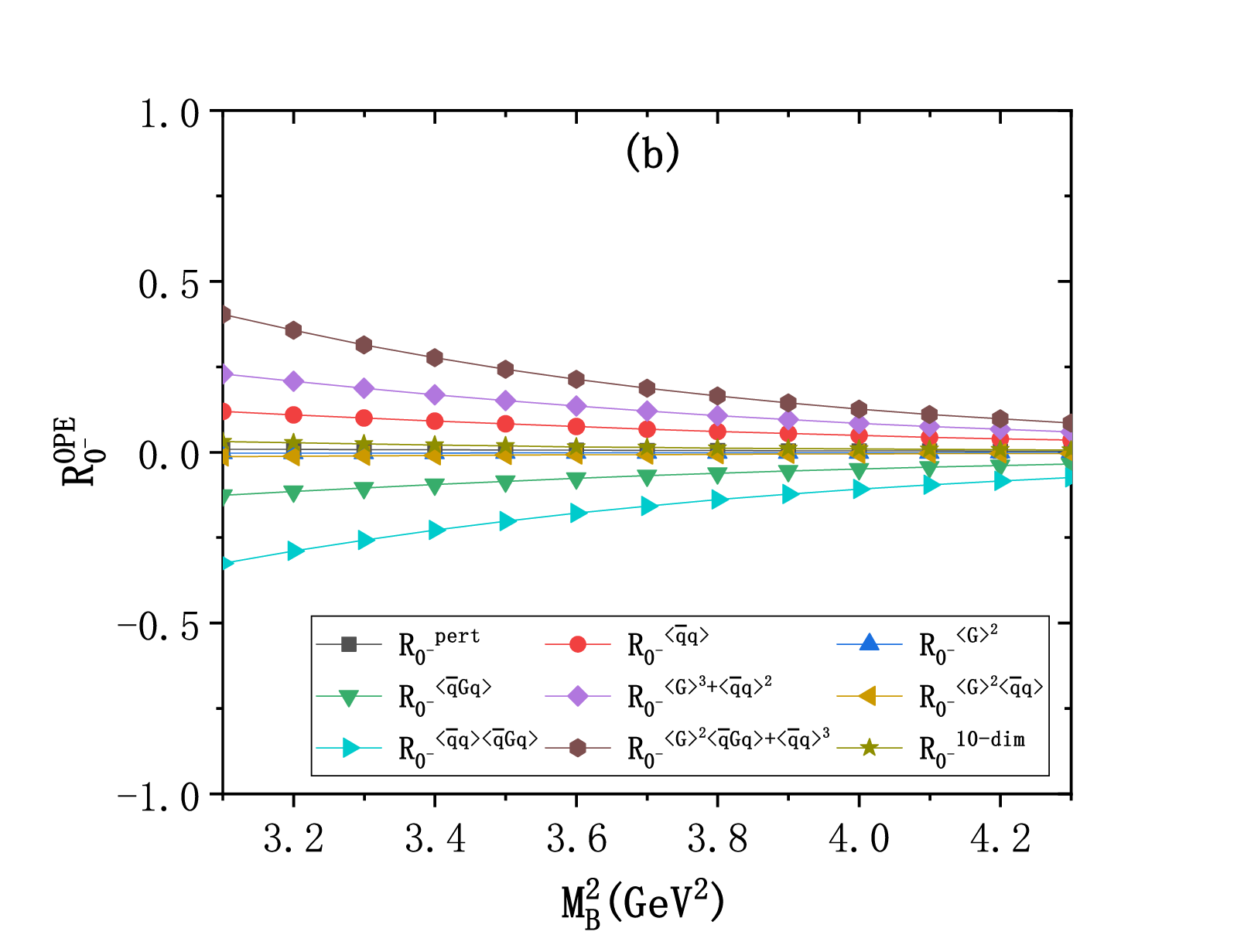}
  \caption{The same caption as in Fig.~\ref{Feyn1}, but for current $j_{6}$.}
  \label{Feyn11}
\end{figure}

\begin{figure}[!htbp]
  \centering
  \includegraphics[width=0.5\textwidth]{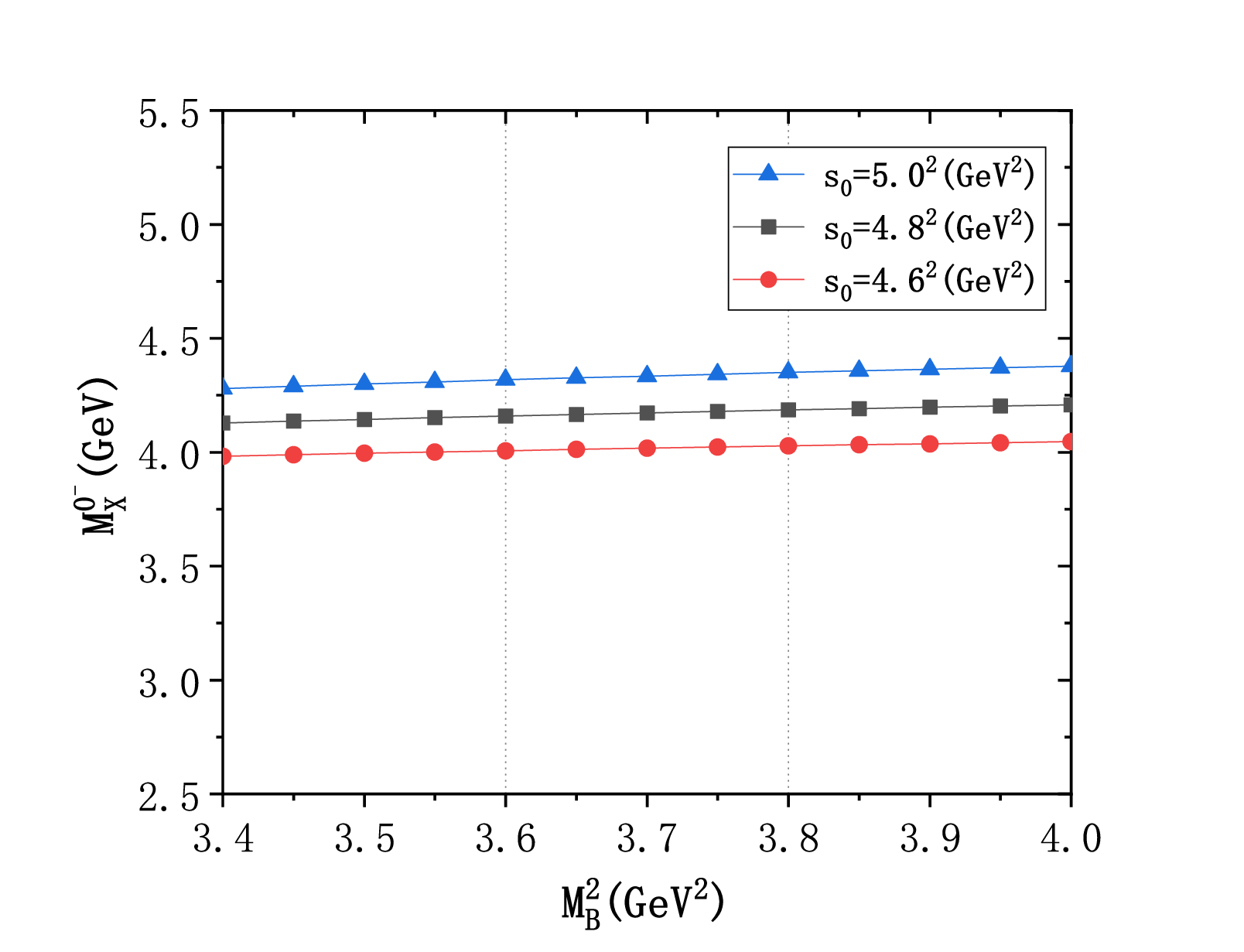}
  \caption{The same caption as in Fig.~\ref{Feyn2}, but for current $j_{6}$.}
  \label{Feyn12}
\end{figure}

For the \(J^{P}=0^{-}\) hexaquark channel interpolated by the current \(j_{1}(x)\), we display in Figs.~\ref{Feyn1}(a) and \ref{Feyn1}(b) the pole contribution \(R_{0^{-}}^{\rm PC}\) and the OPE convergence ratio \(R_{0^{-}}^{\rm cond}\), respectively, as functions of the Borel parameter \(M_{B}^{2}\), with the continuum threshold fixed at \(s_{0}=4.80^2\,{\rm GeV}^{2}\). In Fig.~\ref{Feyn2}, the extracted mass is shown as a function of \(M_{B}^{2}\) for \(s_{0}=4.60^2\,{\rm GeV}^{2}\), \(4.80^2\,{\rm GeV}^{2}\), and \(5.00^2\,{\rm GeV}^{2}\). An analogous analysis is performed for the other five hexaquark interpolating currents. The corresponding results for the interpolating currents \(j_{2}\) through \(j_{6}\) are shown in Figs.~\ref{Feyn3}--\ref{Feyn12}.

Following the above-mentioned criteria of QCD sum rules, we summarize in Table~\ref{tab1} the predicted masses of the hexaquark states, together with the corresponding Borel windows, continuum thresholds \(s_0\), and pole contributions. The results show that, within the optimized Borel windows, the pole contribution satisfies the pole-dominance requirement, while the extracted mass exhibits a stable plateau with respect to the Borel parameter \(M_B^2\). The central values listed in Table~\ref{tab1} are obtained from the most stable regions of the \(M_B^2\) dependence. The quoted uncertainties arise from the variations of the condensate parameters, quark masses, continuum thresholds \(s_0\), and the allowed Borel windows.

\begin{table}[!htb]
\centering
\renewcommand{\arraystretch}{1.20}
\setlength{\tabcolsep}{8pt}
\setlength{\arrayrulewidth}{0.6pt}
\begin{tabular*}{0.88\textwidth}{|@{\extracolsep{\fill}}c|c|c|c|c|}
\hline\hline
Current 
& \(M_B^2~(\mathrm{GeV}^2)\) 
& \(\sqrt{s_{0}}~(\mathrm{GeV})\) 
& PC 
& \(M_{X}~(\mathrm{GeV})\) \\
\hline
\(j_{1}\) & \(3.10\)--\(3.30\) & \(4.80 \pm 0.20\) & \(20\%\)--\(15\%\) & \(4.13_{-0.19}^{+0.19}\) \\
\hline
\(j_{2}\) & \(3.10\)--\(3.30\) & \(4.80 \pm 0.20\) & \(21\%\)--\(15\%\) & \(4.14_{-0.18}^{+0.18}\) \\
\hline
\(j_{3}\) & \(3.30\)--\(3.50\) & \(4.80 \pm 0.20\) & \(21\%\)--\(16\%\) & \(4.17_{-0.15}^{+0.16}\) \\
\hline
\(j_{4}\) & \(3.30\)--\(3.50\) & \(4.80 \pm 0.20\) & \(21\%\)--\(16\%\) & \(4.18_{-0.16}^{+0.16}\) \\
\hline
\(j_{5}\) & \(3.60\)--\(3.80\) & \(4.80 \pm 0.20\) & \(20\%\)--\(15\%\) & \(4.24_{-0.18}^{+0.17}\) \\
\hline
\(j_{6}\) & \(3.60\)--\(3.80\) & \(4.80 \pm 0.20\) & \(20\%\)--\(15\%\) & \(4.24_{-0.18}^{+0.17}\) \\
\hline\hline
\end{tabular*}
\caption{Borel windows \(M_B^2\), continuum thresholds \(\sqrt{s_0}\), pole contributions (PCs), and extracted masses \(M_X\) for the hexaquark states interpolated by the currents \(j_1\)--\(j_6\).}
\label{tab1}
\end{table}

\FloatBarrier

\section{Decay analysis}
\label{sec:decay}

The hexaquark state studied in this work has the same valence quark content as the \(D^{+}D^{-}K^{+}\) system and carries the quantum numbers \(J^{P}=0^{-}\). Since the interpolating currents are constructed from hidden-charm color-octet clusters, the state cannot be regarded as a simple product of three color-singlet mesons. Nevertheless, after color rearrangement and hadronization, it can couple to physical color-singlet hadronic final states with the same conserved quantum numbers. In the present work, we do not calculate the decay widths explicitly. Therefore, the following discussion should be understood as a qualitative analysis based on quantum numbers, flavor conservation, and phase space.

The most direct open-charm three-body decay channel is
\begin{equation}
X^{+} \to D^{+}D^{-}K^{+},
\end{equation}
together with the corresponding charge-related channel
\begin{equation}
X^{+} \to D^{+}\bar{D}^{0}K^{0}.
\end{equation}
The thresholds of these \(D D K\) channels are around \(4.23~\mathrm{GeV}\), which lie within the mass region predicted in this work,
\begin{equation}
M_X = 3.94\text{--}4.41~\mathrm{GeV}.
\end{equation}
Therefore, the \(D D K\) channels are phenomenologically relevant for the present hexaquark candidate. If the physical mass of the state is above the \(D D K\) threshold, the \(D D K\) decay is kinematically allowed. In particular, for three pseudoscalar mesons in relative \(S\) waves, the total parity is negative and the total angular momentum can be \(J=0\), which is consistent with the quantum numbers \(J^{P}=0^{-}\) of the initial state. In this case, the state may appear as an enhancement or a resonance-like structure in the \(D D K\) invariant-mass distribution. If the physical mass lies below the \(D D K\) threshold, the on-shell \(D D K\) decay is forbidden or strongly suppressed, while virtual \(D D K\) components may still affect the near-threshold line shape through rescattering effects.

Besides the direct \(D D K\) final states, other open-charm three-body channels with the same conserved flavor quantum numbers are also possible. Typical examples include
\begin{equation}
X^{+} \to D_s^{+}D^{-}\pi^{+},
\qquad
X^{+} \to D_s^{+}\bar{D}^{0}\pi^{0}.
\end{equation}
These channels have lower thresholds than the \(D D K\) channel and can be reached through quark rearrangement. Since all the final-state mesons are color singlets, such decays require the hidden-charm components in the compact hexaquark state to be converted into physical meson configurations through nonperturbative gluon exchanges. The relative strengths of these modes depend on the overlap between the compact hidden-charm wave function and the corresponding hadronic final states, which cannot be determined from the present two-point sum rule analysis alone.

Hidden-charm three-body final states may also provide useful experimental signatures. Possible channels include
\begin{equation}
X^{+} \to J/\psi\,K^{+}\pi^{0},
\qquad
X^{+} \to J/\psi\,K^{0}\pi^{+},
\end{equation}
and
\begin{equation}
X^{+} \to \eta_c\,K^{+}\pi^{0},
\qquad
X^{+} \to \eta_c\,K^{0}\pi^{+}.
\end{equation}
The \(\eta_c K\pi\) final states can match the quantum numbers of the initial \(0^{-}\) state in an \(S\)-wave three-body configuration, since all three final-state mesons are pseudoscalars. For the \(J/\psi K\pi\) final states, the spin of the \(J/\psi\) requires appropriate nonzero orbital angular momenta among the final-state particles in order to form an overall \(J^{P}=0^{-}\) configuration. Compared with open-charm channels, hidden-charm final states may have smaller branching fractions because they involve a different rearrangement of the quark degrees of freedom. However, they are experimentally attractive since final states containing \(J/\psi\) can often be reconstructed with good resolution.

Therefore, the predicted state may be searched for in three-body open-charm channels such as \(D D K\) and \(D_sD\pi\), as well as in hidden-charm channels such as \(J/\psi K\pi\) and \(\eta_c K\pi\). Since the \(D D K\) threshold lies inside the predicted mass interval, experimental analyses should allow for both above-threshold resonance behavior and possible below-threshold or virtual-state effects. Amplitude analyses of multibody heavy-hadron decays and prompt production processes at LHCb and Belle II would be particularly useful for testing the existence of such a hidden-charm hexaquark candidate.

\section{Conclusions}
\label{sec:conclusion}

In this work, we have studied a compact hidden-charm hexaquark configuration with the same quark content as the \(D^{+}D^{-}K^{+}\) system in the framework of QCD sum rules. Unlike the conventional molecular picture of the \(D D K\) system, the configuration considered here is constructed from three color-octet quark--antiquark clusters,
\begin{equation}
8_{[\bar{d}c]} \otimes 8_{[\bar{c}d]} \otimes 8_{[\bar{s}u]},
\end{equation}
which are coupled to an overall color singlet. This provides a complementary description of possible multiquark dynamics near the open-charm three-body threshold and offers a way to investigate short-distance hidden-charm correlations in QCD.

We constructed six independent local interpolating currents with quantum numbers \(J^{P}=0^{-}\) and calculated the corresponding two-point correlation functions. In the operator product expansion, both perturbative contributions and nonperturbative condensates up to dimension ten were included. By requiring OPE convergence, a sufficient pole contribution, and Borel stability, we obtained reliable sum rule windows for all six currents. The extracted masses are found to be in the range
\begin{equation}
M_X = 3.94\text{--}4.41~\mathrm{GeV}.
\end{equation}
The consistency among different currents suggests that they may couple to the same or similar compact hidden-charm hexaquark configurations.

The predicted mass interval overlaps the \(D D K\) threshold region, indicating that open-charm three-body channels may play an important role in the phenomenology of this state. If the physical mass lies above the \(D D K\) threshold, the \(D D K\) mode can be an allowed strong decay channel; otherwise, below-threshold effects may still appear through virtual components and rescattering. Possible decay channels include \(D D K\), \(D_sD\pi\), and hidden-charm modes such as \(J/\psi K\pi\) and \(\eta_c K\pi\). Since the present analysis is based on two-point QCD sum rules, decay widths and branching fractions cannot be determined. Future studies using three-point sum rules or coupled-channel approaches, together with experimental searches in multibody heavy-hadron decays at LHCb and Belle II, will be essential for clarifying the nature of this possible hidden-charm hexaquark state.

%%%%%%%%%%%%%%%%%%%%%%%%%%%%%%%%%%%%%%%%%%%%%%%%%%%%%%%%%%%%%%%%%%%%%%
\vspace{.7cm} {\bf Acknowledgments} \vspace{.3cm}

This work is supported by the Natural Science Foundation of Hebei Province under Grant No. A2023205038.

%%%%%%%%%%%%%%%%%%%%%%%%%%%%%%%%%%%%%%%%%%%%%%%%%%%%%%%%%%%%%%%%%%%%%%%

\newpage

\section*{Appendix: Spectral density for the current \(j_1\)}
\label{app:rho1}

% ---------- Short-hand commands used in this appendix ----------
\providecommand{\Ixy}{\int_{x_i}^{x_f} dx\,\int_{y_i}^{y_f} dy}
\providecommand{\Ix}{\int_{x_i}^{x_f} dx}

\providecommand{\Fxy}{F_{xy}}
\providecommand{\Hx}{H_x}
\providecommand{\Dxy}{\Delta_{xy}}

\providecommand{\qq}{\langle \bar q q\rangle}
\providecommand{\sscond}{\langle \bar s s\rangle}
\providecommand{\qGq}{\langle \bar q G q\rangle}
\providecommand{\sGs}{\langle \bar s G s\rangle}
\providecommand{\GG}{\langle g_s^2 G^2\rangle}
\providecommand{\GGG}{\langle g_s^3 G^3\rangle}
% -------------------------------------------------------------

For completeness, we present in this appendix the spectral density
\(\rho_1^{\rm OPE}(s)\) entering Eq.~\eqref{eq:15} for the current
\(j_1\) defined in Eq.~\eqref{eq:2}. This expression is given as a representative example; the spectral densities for the other currents have analogous structures and are not displayed explicitly for brevity. The full spectral density is decomposed as
\begin{equation}
\rho_1^{\rm OPE}(s)
=
\rho_1^{\rm pert}(s)
+\rho_1^{(3)}(s)
+\rho_1^{(4)}(s)
+\cdots
+\rho_1^{(10)}(s).
\end{equation}

The short-hand notations used below are defined as
\begin{align}
\Fxy &\equiv m_Q^2(x+y)-xys,
&
\Hx &\equiv m_Q^2x-x(1-x)s,
&
\Dxy &\equiv x+y-1 .
\end{align}
The integration limits are given by
\begin{align}
x_i &= \frac{1-\sqrt{1-4m_Q^2/s}}{2},
&
x_f &= \frac{1+\sqrt{1-4m_Q^2/s}}{2},
\\
y_i &= \frac{m_Q^2x}{-m_Q^2+sx},
&
y_f &= 1-x .
\end{align}
Here \(m_Q=m_c\) for the hidden-charm hexaquark system considered in this work.

The perturbative contribution is
\begin{align}
\rho_{1}^{\rm pert}(s) = -\frac{1}{2^{21} \times 315 \times \pi^{10}} \,
\Ixy \,
\frac{\Fxy^7 \Dxy^4}{x^6 y^6}.
\end{align}

The dimension-three condensate contribution is
\begin{align}
\rho_{1}^{(3)}(s)
={}&
-\frac{\sscond}{2^{17} \times 15 \times \pi^{8}}
\Ixy
\frac{\Fxy^5 m_s \Dxy^2}{x^4 y^4}
\notag\\
&-\frac{\qq}{2^{16} \times 45 \times \pi^{8}}
\Ixy
\frac{\Fxy^5 \Dxy^2}{x^4 y^4}
\notag\\
&\hspace{1.0cm}\times
\Big[
-3m_sxy
+m_Q\bigl(x^2+(y-1)y+x(2y-1)\bigr)
\Big].
\end{align}

The dimension-four condensate contribution is
\begin{align}
\rho_{1}^{(4)}(s)
=
-\frac{\GG}{2^{23} \times 27 \times \pi^{10}}
\Ixy
\frac{\Fxy^4 m_Q^2 \Dxy^4}{x^6 y^6}.
\end{align}

The dimension-five condensate contribution is
\begin{align}
\rho_{1}^{(5)}(s)
={}&
-\frac{\sGs}{2^{17} \times 9 \times \pi^8}
\Ixy
\frac{\Fxy^4 m_s \Dxy}{x^3y^3}
\notag\\
&+
\frac{\qGq}{2^{18} \times 3 \times \pi^8}
\Ixy
\frac{\Fxy^4 \Dxy}{x^4y^4}
\notag\\
&\hspace{1.0cm}\times
\Big[
-2m_sxy
+m_Q\bigl(x^2+(y-1)y+x(2y-1)\bigr)
\Big]
\notag\\
&-
\frac{\qGq}{2^{20} \times 9 \times \pi^8}
\Ixy
\frac{\Fxy^4 g_s m_Q \Dxy^3 (x^2+y^2)}{x^5y^5}
\notag\\
&+
\frac{\sGs}{2^{20} \times 3 \times \pi^8}
\Ixy
\frac{\Fxy^4 g_s m_s \Dxy}{x^3y^3}.
\end{align}

The dimension-six condensate contribution is
\begin{align}
\rho_{1}^{(6)}(s)
={}&
-\frac{\qq^2}{2^{12} \times 9 \times \pi^6}
\Ixy
\frac{\Fxy^3 m_Q \Dxy}{x^3y^3}
\Big[
m_Q\Dxy-2m_s(x+y)
\Big]
\notag\\
&-
\frac{\qq\,\sscond}{2^{14} \times 135 \times \pi^6}
\Ixy
\frac{\Fxy^5 \Dxy^3}{x^4y^4}
\Big[
\Fxy+3m_Qm_s(x+y)
\Big]
\notag\\
&-
\frac{\GG}{2^{25} \times 27 \times \pi^{10}}
\Ixy
\frac{\Fxy^3 \Dxy^4}{x^6y^6}
\Big[
\Fxy(x^3+y^3)+8m_Q^2(x^4+y^4)
\Big].
\end{align}

The dimension-seven condensate contribution is
\begin{align}
\rho_{1}^{(7)}(s)
={}&
-\frac{\sscond\,\GG}{2^{18} \times 9 \times \pi^8}
\Ixy
\frac{\Fxy^2 m_Q^2 m_s \Dxy^2 (x^3+y^3)}{x^4y^4}
\notag\\
&-
\frac{\qq\,\GG}{2^{17} \times 27 \times \pi^8}
\Ixy
\frac{\Fxy^2 m_Q \Dxy^2 (x^3+y^3)}{x^5y^5}
\notag\\
&\hspace{1.0cm}\times
\Big\{
\Fxy\Dxy
+m_Q
\Big[
-3m_sxy
+m_Q\bigl(x^2+(y-1)y+x(2y-1)\bigr)
\Big]
\Big\}.
\end{align}

The dimension-eight condensate contribution is
\begin{align}
\rho_{1}^{(8)}(s)
={}&
\frac{\sscond\,\qGq}{2^{14} \times 9 \times \pi^6}
\Ixy
\frac{\Fxy^2}{x^2y^2}
\Big[
2\Fxy+3m_Qm_s(x+y)
\Big]
\notag\\
&+
\frac{\qq\,\qGq}{2^{12} \times 3 \times \pi^6}
\Ixy
\frac{\Fxy^2 m_Q}{x^2y^2}
\Big[
m_Q\Dxy-m_s(x+y)
\Big]
\notag\\
&+
\frac{\qq\,\sGs}{2^{15} \times 135 \times \pi^6}
\Ixy
\frac{\Fxy^4 \Dxy^2}{x^3y^3}
\Big[
3\Fxy+5m_Qm_s(x+y)
\Big].
\end{align}

The dimension-nine condensate contribution is
\begin{align}
\rho_{1}^{(9)}(s)
={}&
\frac{\qq^3}{2^{8} \times 9 \times \pi^4}
\Ixy
\frac{\Fxy m_Q^2 m_s}{xy}
\notag\\
&-
\frac{\qq^2\sscond}{2^{9} \times 9 \times \pi^4}
\Ixy
\frac{\Fxy m_Q}{x^2y^2}
\Big[
m_Qm_sxy+\Fxy(x+y)
\Big]
\notag\\
&+
\frac{\GG\,\qGq}{2^{19} \times 9 \times \pi^8}
\Ixy
\frac{\Fxy m_Q \Dxy (x^3+y^3)}{x^4y^4}
\notag\\
&\hspace{1.0cm}\times
\Big\{
3\Fxy\Dxy
+2m_Q
\Big[
-2m_sxy
+m_Q\bigl(x^2+(y-1)y+x(2y-1)\bigr)
\Big]
\Big\}
\notag\\
&+
\frac{\GG\,\sGs}{2^{17} \times 27 \times \pi^8}
\Ixy
\frac{\Fxy m_Q^2 m_s \Dxy (x^3+y^3)}{x^3y^3}
\notag\\
&-
\frac{\GGG\,\qq}{2^{19} \times 27 \times \pi^8}
\Ixy
\frac{\Fxy \Dxy^2}{x^5y^5}
\notag\\
&\hspace{1.0cm}\times
\Bigg\{
4m_Q^2(x^4+y^4)
\Big[
-3m_sxy
+m_Q\bigl(x^2+(y-1)y+x(2y-1)\bigr)
\Big]
\notag\\
&\hspace{1.0cm}
+\Fxy
\Big[
-3m_sxy(x^3+y^3)
\notag\\
&\hspace{1.8cm}
+m_Q
\Big(
6x^5+x^3(y-1)y+x^2y^3+6(y-1)y^4
\notag\\
&\hspace{2.8cm}
+x^4(-6+7y)+xy^3(-1+7y)
\Big)
\Big]
\Bigg\}
\notag\\
&-
\frac{\GGG\,\sscond}{2^{20} \times 9 \times \pi^8}
\Ixy
\frac{\Fxy m_s \Dxy^2}{x^4y^4}
\Big[
\Fxy(x^3+y^3)+4m_Q^2(x^4+y^4)
\Big].
\end{align}

The dimension-ten condensate contribution is
\begin{align}
\rho_{1}^{(10)}(s)
={}&
-\frac{\qGq^2}{2^{14} \times 3 \times \pi^6}
\Ixy
\frac{\Fxy m_Q^2}{xy}
\notag\\
&-
\frac{\GG\,\qq^2}{2^{14} \times 27 \times \pi^6}
\Ixy
\frac{m_Q \Dxy}{x^3y^3}
\notag\\
&\hspace{1.0cm}\times
\Bigg\{
m_Q^2(x^3+y^3)
\Big[
m_Q\Dxy-2m_s(x+y)
\Big]
\notag\\
&\hspace{1.0cm}
+3\Fxy
\Big[
m_Q\bigl(x^3+x^2(y-1)+xy^2+(y-1)y^2\bigr)
-2m_s(x^3+y^3)
\Big]
\Bigg\}
\notag\\
&-
\frac{\GG\,\qq\,\sscond}{2^{14} \times 27 \times \pi^6}
\Ixy
\frac{m_Q \Dxy (x^3+y^3)}{x^3y^3}
\Big[
\Fxy(2m_Q+3m_s)+m_Q^2m_s(x+y)
\Big]
\notag\\
&+
\frac{\qGq^2}{2^{14} \times 3 \times \pi^6}
\Ix
\frac{\Hx m_Qm_s}{(x-1)x}
\notag\\
&-
\frac{\qGq\,\sGs}{2^{15} \times 9 \times \pi^6}
\Ix
\frac{\Hx(3\Hx+2m_Qm_s)}{(x-1)x}.
\end{align}


\begin{thebibliography}{99}

\bibitem{Gell-Mann:1964ewy}
M.~Gell-Mann,
``A Schematic Model of Baryons and Mesons,''
Phys. Lett. \textbf{8}, 214--215 (1964)
\doi{10.1016/S0031-9163(64)92001-3}.

\bibitem{Zweig:1964ruk}
G.~Zweig,
``An SU(3) model for strong interaction symmetry and its breaking. Version 1,''
CERN-TH-401.

\bibitem{Jaffe:1976yi}
R.~L.~Jaffe,
``Perhaps a Stable Dihyperon,''
Phys. Rev. Lett. \textbf{38}, 195--198 (1977)
[erratum: Phys. Rev. Lett. \textbf{38}, 617 (1977)]
\doi{10.1103/PhysRevLett.38.195}.

\bibitem{Lebed:2016hpi}
R.~F.~Lebed, R.~E.~Mitchell and E.~S.~Swanson,
``Heavy-Quark QCD Exotica,''
Prog. Part. Nucl. Phys. \textbf{93}, 143--194 (2017)
\doi{10.1016/j.ppnp.2016.11.003}
[arXiv:1610.04528 [hep-ph]].

\bibitem{Chen:2016qju}
H.~X.~Chen, W.~Chen, X.~Liu and S.~L.~Zhu,
``The hidden-charm pentaquark and tetraquark states,''
Phys. Rept. \textbf{639}, 1--121 (2016)
\doi{10.1016/j.physrep.2016.05.004}
[arXiv:1601.02092 [hep-ph]].

\bibitem{WASA-at-COSY:2014dmv}
P.~Adlarson \textit{et al.} [WASA-at-COSY],
``Evidence for a New Resonance from Polarized Neutron-Proton Scattering,''
Phys. Rev. Lett. \textbf{112}, 202301 (2014)
\doi{10.1103/PhysRevLett.112.202301}
[arXiv:1402.6844 [nucl-ex]].

\bibitem{Belle:2003nnu}
S.~K.~Choi \textit{et al.} [Belle],
``Observation of a narrow charmonium-like state in exclusive $B^\pm \to K^\pm \pi^+ \pi^- J/\psi$ decays,''
Phys. Rev. Lett. \textbf{91}, 262001 (2003)
\doi{10.1103/PhysRevLett.91.262001}
[arXiv:hep-ex/0309032 [hep-ex]].

\bibitem{LHCb:2016lve}
R.~Aaij \textit{et al.} [LHCb],
``Evidence for exotic hadron contributions to $\Lambda_b^0 \to J/\psi p \pi^-$ decays,''
Phys. Rev. Lett. \textbf{117}, 082003 (2016)
\doi{10.1103/PhysRevLett.117.082003}
[arXiv:1606.06999 [hep-ex]].

\bibitem{LHCb:2019kea}
R.~Aaij \textit{et al.} [LHCb],
``Observation of a narrow pentaquark state, $P_c(4312)^+$, and of two-peak structure of the $P_c(4450)^+$,''
Phys. Rev. Lett. \textbf{122}, 222001 (2019)
\doi{10.1103/PhysRevLett.122.222001}
[arXiv:1904.03947 [hep-ex]].

\bibitem{Ali:2017jda}
A.~Ali, J.~S.~Lange and S.~Stone,
``Exotics: Heavy Pentaquarks and Tetraquarks,''
Prog. Part. Nucl. Phys. \textbf{97}, 123--198 (2017)
\doi{10.1016/j.ppnp.2017.08.003}
[arXiv:1706.00610 [hep-ph]].

\bibitem{Esposito:2016noz}
A.~Esposito, A.~Pilloni and A.~D.~Polosa,
``Multiquark Resonances,''
Phys. Rept. \textbf{668}, 1--97 (2017)
\doi{10.1016/j.physrep.2016.11.002}
[arXiv:1611.07920 [hep-ph]].

\bibitem{Richard:2016eis}
J.~M.~Richard,
``Exotic hadrons: review and perspectives,''
Few Body Syst. \textbf{57}, 1185--1212 (2016)
\doi{10.1007/s00601-016-1159-0}
[arXiv:1606.08593 [hep-ph]].

\bibitem{Liu:2019zoy}
Y.~R.~Liu, H.~X.~Chen, W.~Chen, X.~Liu and S.~L.~Zhu,
``Pentaquark and Tetraquark states,''
Prog. Part. Nucl. Phys. \textbf{107}, 237--320 (2019)
\doi{10.1016/j.ppnp.2019.04.003}
[arXiv:1903.11976 [hep-ph]].

\bibitem{BESIII:2013ris}
M.~Ablikim \textit{et al.} [BESIII],
``Observation of a Charged Charmoniumlike Structure in $e^+e^- \to \pi^+\pi^- J/\psi$ at $\sqrt{s}=4.26$ GeV,''
Phys. Rev. Lett. \textbf{110}, 252001 (2013)
\doi{10.1103/PhysRevLett.110.252001}
[arXiv:1303.5949 [hep-ex]].

\bibitem{Brambilla:2019esw}
N.~Brambilla, S.~Eidelman, C.~Hanhart, A.~Nefediev, C.~P.~Shen, C.~E.~Thomas, A.~Vairo and C.~Z.~Yuan,
``The $XYZ$ states: experimental and theoretical status and perspectives,''
Phys. Rept. \textbf{873}, 1--154 (2020)
\doi{10.1016/j.physrep.2020.05.001}
[arXiv:1907.07583 [hep-ex]].

\bibitem{Wang:2019tlw}
Z.~G.~Wang,
``Analysis of the hidden-charm tetraquark mass spectrum with the QCD sum rules,''
Phys. Rev. D \textbf{102}, 014018 (2020)
\doi{10.1103/PhysRevD.102.014018}
[arXiv:1908.07914 [hep-ph]].

\bibitem{Guo:2017jvc}
F.~K.~Guo, C.~Hanhart, U.~G.~Mei{\ss}ner, Q.~Wang, Q.~Zhao and B.~S.~Zou,
``Hadronic molecules,''
Rev. Mod. Phys. \textbf{90}, 015004 (2018)
\doi{10.1103/RevModPhys.90.015004}
[arXiv:1705.00141 [hep-ph]].

\bibitem{Voloshin:2004mh}
M.~B.~Voloshin,
``Heavy quark spin selection rule and the properties of the X(3872),''
Phys. Lett. B \textbf{604}, 69--73 (2004)
\doi{10.1016/j.physletb.2004.11.003}
[arXiv:hep-ph/0408321 [hep-ph]].

\bibitem{Tornqvist:2004qy}
N.~A.~Tornqvist,
``Isospin breaking of the narrow charmonium state of Belle at 3872 MeV as a deuson,''
Phys. Lett. B \textbf{590}, 209--215 (2004)
\doi{10.1016/j.physletb.2004.03.077}
[arXiv:hep-ph/0402237 [hep-ph]].

\bibitem{Maiani:2004vq}
L.~Maiani, F.~Piccinini, A.~D.~Polosa and V.~Riquer,
``Diquark-antidiquarks with hidden or open charm and the nature of X(3872),''
Phys. Rev. D \textbf{71}, 014028 (2005)
\doi{10.1103/PhysRevD.71.014028}
[arXiv:hep-ph/0412098 [hep-ph]].

\bibitem{Wang:2013vex}
Z.~G.~Wang and T.~Huang,
``Analysis of the $X(3872)$, $Z_c(3900)$ and $Z_c(3885)$ as axial-vector tetraquark states with QCD sum rules,''
Phys. Rev. D \textbf{89}, 054019 (2014)
\doi{10.1103/PhysRevD.89.054019}
[arXiv:1310.2422 [hep-ph]].

\bibitem{Guo:2015umn}
F.~K.~Guo, U.~G.~Mei{\ss}ner, W.~Wang and Z.~Yang,
``How to reveal the exotic nature of the P$_c$(4450),''
Phys. Rev. D \textbf{92}, 071502 (2015)
\doi{10.1103/PhysRevD.92.071502}
[arXiv:1507.04950 [hep-ph]].

\bibitem{Szczepaniak:2015eza}
A.~P.~Szczepaniak,
``Triangle Singularities and XYZ Quarkonium Peaks,''
Phys. Lett. B \textbf{747}, 410--416 (2015)
\doi{10.1016/j.physletb.2015.06.029}
[arXiv:1501.01691 [hep-ph]].

%\cite{Wang:2017sto}
\bibitem{Wang:2017sto}
  Z.~G.~Wang,
  ``Analysis of the scalar doubly charmed hexaquark state with QCD sum rules,''
  Eur.\ Phys.\ J.\ C \textbf{77}, 642 (2017),
  \href{https://doi.org/10.1140/epjc/s10052-017-5207-9}{doi:10.1140/epjc/s10052-017-5207-9}
  [arXiv:1707.09767 [hep-ph]].

%\cite{Wang:2021qmn}
\bibitem{Wang:2021qmn}
  X.~W.~Wang, Z.~G.~Wang, and G.~L.~Yu,
  ``Study of $\Lambda_c\Lambda_c$ dibaryon and $\Lambda_c\bar{\Lambda}_c$
  baryonium states via QCD sum rules,''
  Eur.\ Phys.\ J.\ A \textbf{57}, 275 (2021),
  \href{https://doi.org/10.1140/epja/s10050-021-00576-8}{doi:10.1140/epja/s10050-021-00576-8}
  [arXiv:2107.04751 [hep-ph]].

%\cite{Wan:2019ake}
\bibitem{Wan:2019ake}
  B.~D.~Wan, L.~Tang, and C.~F.~Qiao,
  ``Hidden-bottom and -charm hexaquark states in QCD sum rules,''
  Eur.\ Phys.\ J.\ C \textbf{80}, 121 (2020),
  \href{https://doi.org/10.1140/epjc/s10052-020-7701-8}{doi:10.1140/epjc/s10052-020-7701-8}
  [arXiv:1912.12046 [hep-ph]].

%\cite{Zhang:2025jqx}
\bibitem{Zhang:2025jqx}
  X.~H.~Zhang and C.~F.~Qiao,
  ``Mass spectra of $\Lambda_Q\bar{\Sigma}_Q$ hexaquark states in QCD sum
  rules,''
  arXiv:2512.22019 [hep-ph].

%\cite{Wan:2023epq}
\bibitem{Wan:2023epq}
  B.~D.~Wan,
  ``Mass spectra of $0^{--}$ and $0^{+-}$ hidden-heavy baryoniums,''
  Eur.\ Phys.\ J.\ C \textbf{84}, 760 (2024),
  \href{https://doi.org/10.1140/epjc/s10052-024-13126-5}{doi:10.1140/epjc/s10052-024-13126-5}
  [arXiv:2311.13161 [hep-ph]].

%\cite{Wang:2026lta}
\bibitem{Wang:2026lta}
  X.~W.~Wang, Z.~G.~Wang, and G.~L.~Yu,
  ``Search for the $\Lambda_c\Sigma_c$ and $\bar{\Lambda}_c\Sigma_c$ dibaryon
  structures via the QCD sum rules,''
  arXiv:2604.14597 [hep-ph].

%\cite{Chen:2016ymy}
\bibitem{Chen:2016ymy}
  H.~X.~Chen, D.~Zhou, W.~Chen, X.~Liu, and S.~L.~Zhu,
  ``Searching for hidden-charm baryonium signals in QCD sum rules,''
  Eur.\ Phys.\ J.\ C \textbf{76}, 602 (2016),
  \href{https://doi.org/10.1140/epjc/s10052-016-4459-0}{doi:10.1140/epjc/s10052-016-4459-0}
  [arXiv:1605.07453 [hep-ph]].
  
  %\cite{Wang:2021pua}
\bibitem{Wang:2021pua}
  X.~W.~Wang and Z.~G.~Wang,
  ``Search for the charmed baryonium and dibaryon structures via the QCD sum
  rules,''
  Adv.\ High Energy Phys.\ \textbf{2022}, 6224597 (2022),
  \href{https://doi.org/10.1155/2022/6224597}{doi:10.1155/2022/6224597}
  [arXiv:2110.14133 [hep-ph]].
  
\bibitem{MartinezTorres:2019}
A.~Martinez Torres, K.~P.~Khemchandani and L.~S.~Geng,
``Bound state formation in the $DDK$ system,''
Phys. Rev. D \textbf{99}, 076017 (2019)
\doi{10.1103/PhysRevD.99.076017}
[arXiv:1809.01059 [hep-ph]].

\bibitem{PhysRevD.103.L031501}
T.~W.~Wu, M.~Z.~Liu and L.~S.~Geng,
``Excited $K$ meson, $K_c(4180)$, with hidden charm as a $\bar D \bar D K$ bound state,''
Phys. Rev. D \textbf{103}, L031501 (2021)
\doi{10.1103/PhysRevD.103.L031501}
[arXiv:2102.08109 [hep-ph]].

\bibitem{Wu:2019vsy}
T.~W.~Wu, M.~Z.~Liu, L.~S.~Geng, E.~Hiyama and M.~P.~Valderrama,
``$DK$, $DDK$, and $DDDK$ molecules---understanding the nature of the $D_{s0}^*(2317)$,''
Phys. Rev. D \textbf{100}, 034029 (2019)
\doi{10.1103/PhysRevD.100.034029}
[arXiv:1906.11995 [hep-ph]].

\bibitem{Wu:2025fzx}
T.~W.~Wu, M.~Z.~Liu and L.~S.~Geng,
``Implication of the Existence of $J^{PC}=0^{--}\,\bar D_s DK$ Bound State on the Nature of $D_{s0}^*(2317)$, and a New Configuration of Exotic State,''
Phys. Rev. Lett. \textbf{135}, 031902 (2025)
\doi{10.1103/3gbb-ms8c}
[arXiv:2501.11358 [hep-ph]].

\bibitem{Huang:2019qmw}
Y.~Huang, M.~Z.~Liu, Y.~W.~Pan, L.~S.~Geng, A.~Mart{\'\i}nez Torres and K.~P.~Khemchandani,
``Strong decays of the explicitly exotic doubly charmed $DDK$ bound state,''
Phys. Rev. D \textbf{101}, 014022 (2020)
\doi{10.1103/PhysRevD.101.014022}
[arXiv:1909.09021 [hep-ph]].

\bibitem{Zhang:2024yfj}
Z.~Zhang, X.~Y.~Hu, G.~He, J.~Liu, J.~A.~Shi, B.~N.~Lu and Q.~Wang,
``Binding of the three-hadron $DD^*K$ system from the lattice effective field theory,''
Phys. Rev. D \textbf{111}, 036002 (2025)
\doi{10.1103/PhysRevD.111.036002}
[arXiv:2409.01325 [hep-ph]].

\bibitem{Shifman:1978bx}
M.~A.~Shifman, A.~I.~Vainshtein and V.~I.~Zakharov,
``QCD and Resonance Physics. Theoretical Foundations,''
Nucl. Phys. B \textbf{147}, 385--447 (1979)
\doi{10.1016/0550-3213(79)90022-1}.

\bibitem{Shifman:1978by}
M.~A.~Shifman, A.~I.~Vainshtein and V.~I.~Zakharov,
``QCD and Resonance Physics: Applications,''
Nucl. Phys. B \textbf{147}, 448--518 (1979)
\doi{10.1016/0550-3213(79)90023-3}.

\bibitem{Reinders:1984sr}
L.~J.~Reinders, H.~Rubinstein and S.~Yazaki,
``Hadron Properties from QCD Sum Rules,''
Phys. Rept. \textbf{127}, 1 (1985)
\doi{10.1016/0370-1573(85)90065-1}.

\bibitem{Albuquerque:2013ija}
R.~M.~Albuquerque,
``Charmonium Exotic States,''
arXiv:1306.4671 [hep-ph].

\bibitem{Govaerts:1984hc}
J.~Govaerts, L.~J.~Reinders, H.~R.~Rubinstein and J.~Weyers,
``Hybrid Quarkonia From QCD Sum Rules,''
Nucl. Phys. B \textbf{258}, 215--229 (1985)
\doi{10.1016/0550-3213(85)90609-1}.

\bibitem{P.Col}
P.~Colangelo and A.~Khodjamirian,
``QCD Sum Rules, a Modern Perspective,''
in {\it At the Frontier of Particle Physics / Handbook of QCD},
edited by M.~Shifman (World Scientific, Singapore, 2001), Vol.~3, pp.~1495--1576
\doi{10.1142/9789812810458_0033}
[arXiv:hep-ph/0010175].

\bibitem{Colangelo:2000dp}
P.~Colangelo and A.~Khodjamirian,
``QCD sum rules, a modern perspective,''
arXiv:hep-ph/0010175 [hep-ph].

\bibitem{Narison:1989aq}
S.~Narison,
``QCD spectral sum rules,''
World Sci. Lect. Notes Phys. \textbf{26}, 1--527 (1989)
\doi{10.1142/9789812799265}.

\bibitem{ParticleDataGroup:2022pth}
R.~L.~Workman \textit{et al.} [Particle Data Group],
``Review of Particle Physics,''
PTEP \textbf{2022}, 083C01 (2022)
\doi{10.1093/ptep/ptac097}.

\bibitem{Zhang:2025qmg}
S.~Q.~Zhang and C.~F.~Qiao,
``Baryons and baryoniums in the perspective of QCD sum rules,''
arXiv:2512.24706 [hep-ph].

\bibitem{Finazzo:2011he}
S.~I.~Finazzo, M.~Nielsen and X.~Liu,
``QCD sum rule calculation for the charmonium-like structures in the $J/\psi \phi$ and $J/\psi \omega$ invariant mass spectra,''
Phys. Lett. B \textbf{701}, 101--106 (2011)
\doi{10.1016/j.physletb.2011.05.042}
[arXiv:1102.2347 [hep-ph]].

\bibitem{Qiao:2013raa}
C.~F.~Qiao and L.~Tang,
``Estimating the mass of the hidden charm $1^+(1^+)$ tetraquark state via QCD sum rules,''
Eur. Phys. J. C \textbf{74}, 3122 (2014)
\doi{10.1140/epjc/s10052-014-3122-x}
[arXiv:1307.6654 [hep-ph]].

\bibitem{Qiao:2013dda}
C.~F.~Qiao and L.~Tang,
``Interpretation of $Z_c(4025)$ as the hidden charm tetraquark states via QCD Sum Rules,''
Eur. Phys. J. C \textbf{74}, 2810 (2014)
\doi{10.1140/epjc/s10052-014-2810-x}
[arXiv:1308.3439 [hep-ph]].

\end{thebibliography}
\end{document}